%% file: draft1.tex
\documentclass[a4paper,fleqn,usenatbib]{mnras}

\usepackage[T1]{fontenc}
\usepackage{ae,aecompl}
\usepackage{fixltx2e}

\usepackage{graphicx}	
\usepackage{amsmath}	
\usepackage{amssymb}	

\usepackage{xcolor}		
\usepackage{stackengine}

\input{mycommands.tex}

\title[Slowly-spinning SMBHs in the SDSS]{Testing the Completeness of the SDSS Colour Selection for Ultramassive, Slowly Spinning Black Holes} 

\author[Bertemes et al.]{Caroline Bertemes,$^{1}$\thanks{E-mail: cberteme@phys.ethz.ch} 
Benny Trakhtenbrot,$^{1}$\thanks{Zwicky fellow. E-mail: benny.trakhtenbrot@phys.ethz.ch}
Kevin Schawinski,$^{1}$
Chris Done,$^{2}$
\newauthor {Martin Elvis$^{3}$}\\
$^{1}$ Institute for Astronomy, Department of Physics, ETH Zurich,Wolfgang-Pauli-Strasse 27, CH-8093 Zurich, Switzerland\\ 
$^{2}$ Department of Physics, University of Durham, South Road, Durham DH1 3LE, UK\\
$^{3}$ Harvard-Smithsonian Center for Astrophysics, 60 Garden St., Cambridge, MA 02138, USA\\
}

\date{Accepted XXX. Received YYY; in original form ZZZ}

\pubyear{2016}

\begin{document}
\label{firstpage}
\pagerange{\pageref{firstpage}--\pageref{lastpage}}
\maketitle

\begin{abstract}

We investigate the sensitivity of the colour-based quasar selection algorithm of the Sloan Digital Sky Survey to several key physical parameters of supermassive black holes (SMBHs), focusing on BH spin (\as) at the high BH-mass regime ($\mbh\geqslant10^9\,\Msun$). 
We use a large grid of model spectral energy distribution, assuming geometrically-thin, optically-thick accretion discs, and spanning a wide range of five physical parameters: BH mass \mbh, BH spin \as, Eddington ratio \lledd , redshift $z$, and inclination angle $inc$. 
Based on the expected fluxes in the SDSS imaging \textit{ugriz} bands, we find that $\sim 99.8$\% of our models with $\mbh \leqslant 10^{9.5}\, \Msun$ are selected as quasar candidates and thus would have been targeted for spectroscopic follow-up. 
However, in the extremely high-mass regime, $\geqslant 10^{10} \Msun$, we identify a bias against slowly/retrograde spinning SMBHs. 
The fraction of SEDs that would have been selected as quasar candidates drops below $\sim50\%$ for $\as<0$ across $0.5<z<2$. 
For particularly massive BHs, with $\mbh \simeq 3\times10^{10}\,\Msun$, this rate drops below $\sim20$\%, and can be yet lower for specific redshifts.
We further find that the chances of identifying any hypothetical sources with $\mbh = 10^{11}\,\Msun$ by colour selection would be extremely low at the level of $\sim 3$\%.
Our findings, along with several recent theoretical arguments and empirical findings, demonstrate that the current understanding of the SMBH population at the high-\mbh, and particularly the low- or retrograde-spinning regime, is highly incomplete.
\end{abstract}




\section{Introduction}
\label{sec:Intro}

The spins of astrophysical, supermassive black holes (SMBHs), along with their masses (\mbh) and the growth rates ($\dot{M}_{\rm BH}$), are key parameters in understanding the physics of their close environments and their growth history.
The BH spin\footnotemark affects the space-time metric in the close vicinity, by setting the distance to the innermost stable (circular) orbit (ISCO), which can vary between $\sim$6, 1, and 9 gravitational radii, for a non-rotating, maximally prograde rotating, and maximally retrograde rotating BH, respectively ($\as=0$, $0.998$, and $-1$, respectively). 
This, in turn, may affect the total output and spectral dependence of the radiation emitted from any gas accretion flow, since higher energy photons may be emitted from smaller radii.
Indeed, within the standard geometrically-thin, optically-thick accretion disc model, the radiative efficiency \re\ covers almost an order of magnitude, ranging $\eta\simeq 0.038-0.32$ \cite[see, e.g.,][]{Netzer2013_book}.
This efficiency then determines the efficiency by which the BH accumulates mass, for any given gas accretion rate through the disc.
Moreover, the spin of the BH may evolve due to accretion and coalescence events.
A large number of isotropically oriented accretion events would lead to a ``spin-down'' of the BH, reaching $\as\ltsim0.2$, while prolonged accretion and/or drastically anisotropic accretion events would ``spin up'' the BH, reaching $\as\simeq1$ \cite[e.g.,][]{KingPringle2006, King2008_spindown,Dotti2013}.
The predictions of the two scenarios differ mainly in the extremely high-mass regime,  $\mbh > 10^{9}\, \Msol$, as such massive SMBHs have experienced more accretion episodes than their lower-mass counterparts.
Understanding the demographics of BH spin among both relic and accreting SMBHs (i.e., AGN), and particularly the most massive ones could therefore provide key insights to SMBH growth history, and potentially the galaxy-scale processes that drive it \cite[e.g.,][]{Volonteri2013_spin}.

\footnotetext{Throughout this work, we describe the spin via the dimensionless Kerr parameter $\as = J/\mbh^{2}$ (setting $G=c=1$), where $J$ denotes the angular momentum of the BH and \mbh\ its mass. 
\as\ can then vary between $-1$ and $0.998$ \cite[see][]{Thorne1974}.}

Measuring, or even constraining spins in SMBHs is, however, challenging. 
One common approach is based on the reflection-dominated, gravitationally-broadened component of the K$\alpha$ Iron emission line observed near 6.7 \kev, which probes the disc region close to the ISCO. 
So far, this approach was applied to about 20 low-redshift AGN, providing predominantly high \as\ measurements, with several objects having $\as\geqslant0.99$ \citep{Reynolds2014_Ka_spins}. 
However, these measurements may be affected by systematic uncertainties in modelling the underlying continuum and the physical interpretation of the K$\alpha$ line profile \cite[e.g.,][and references therein]{Miller2013_no_spin}.
Moreover, the limited sensitivity of present-day facilities limits the usage of this method to low redshifts, with only a few exceptions of very bright and/or lensed AGN at $z\sim1$ \cite[][]{Reis2014_spin,Reynolds2014_Einstein_cross}.
An alternative approach, which essentially relies on fitting the (rest-frame) UV-optical spectral energy distributions (SED) of AGN for which \mbh\ is known, has been recently employed for several larger samples of higher-redshift systems, reaching $z\sim3.5$ \cite[e.g.,][]{DavisLaor2011_AD,Wu2013_eff,NetzerTrakht2014_slim,Trakhtenbrot2014_hiz_spin,Capellupo2015_ADs,Capellupo2016_XS_pap3}.
These studies have generally found extremely high BH spins for the most massive BHs, but typically lower spins for the lower mass objects. 
One notable exception, of an extremely massive BH with $\as=0.3$, was  reported by \cite{Czerny2011}.
Indirect arguments involving the ensemble properties of AGN across all redshifts (e.g., \citealt{Elvis2002_eta,YuTremaine2002_soltan_arg}, following the so-called ``Soltan argument'',  \citealt{Soltan1982}) support an efficiency of $\re\sim10-15\%$, corresponding to $\as\sim0.8-0.9$. 
However, such calculations are very sensitive to a series of simplifying assumptions.
For example, the recent revision of the relation between black hole mass and host galaxy mass in the local Universe \cite[][]{KormendyHo2013_MM_Rev} leads to lower efficiency and hence spin, giving either constant efficiency of $\re\simeq0.055$ (consistent with non-spinning BHs) or with \re\ which increases with \mbh, so that the most massive BHs have extremely high spins \cite[e.g.,][]{Ueda2014}.
Thus, most of the available data supports high BH spins amongst the most massive BHs, favouring the "spin-up" scenario. 
The Iron K$\alpha$ studies suggest extreme spins for the lower mass BHs as well.

With both of the aforementioned methods favouring high spin values, it is crucial to investigate the role of any selection effects and/or biases that could be affecting these analyses, particularly in the high-mass regime ($\mbh \gtrsim 10^{9}\,\Msun$). 
It has already been argued that AGN selection which is based on high-ionization emission lines would be biased against BH with high \mbh\ and low (or retrograde) spins, since the lack of UV photons in their SEDs would lead to a low fraction of ionizing photons (see 13.6 eV line in figure \ref{fig:DL_reproduced}), and thus to the observation of weak lined quasars \cite[WLQs; see, e.g.,][]{LaorDavis2011_WLQs}.
Another concern is that the prescriptions used in the continuum-fitting approach would 
lead to bias towards high \re, and therefore high BH spins, when applied to large (optically) flux-limited  surveys \cite[such as the SDSS;][]{Raimundo2012}.
In the present work, we will show that on top of this, the colour selection in such surveys may also be favouring high-spins among high-\mbh\ systems.

The outline of the paper is as follows. 
In Section~\ref{sec:calculations}, we describe our models and the related assumptions and briefly outline the SDSS colour-based target selection algorithm. 
In Section~\ref{sec:ResultsDiscussion}, we present and discuss our main results regarding the ability of the SDSS target selection algorithm to identify the calculated thin accretion disc SEDs. 
We summarize our main findings in Section~\ref{sec:Summary}.
Throughout this work we convert luminosities to fluxes and to magnitudes assuming a cosmological model with $\Omega_{\Lambda}=0.7$, $\Omega_{\rm M}=0.3$, and
$H_{0}=70\,\kms\,{\rm Mpc}^{-1}$, and the AB magnitude system.

\section{Calculations}
\label{sec:calculations}

Our analysis focuses on testing the completeness of the colour-colour quasar selection algorithm of the Sloan Digital Sky Survey \cite[SDSS,][]{York2000, Richards2002} for SMBHs properties, and particularly BH spin (\as).
This is done by using a large grid of thin accretion disc (AD) models, calculated for different physical properties of the accreting SMBH.

\subsection{Model spectra}

\begin{figure*}
\centering
\includegraphics[width=0.49\textwidth]{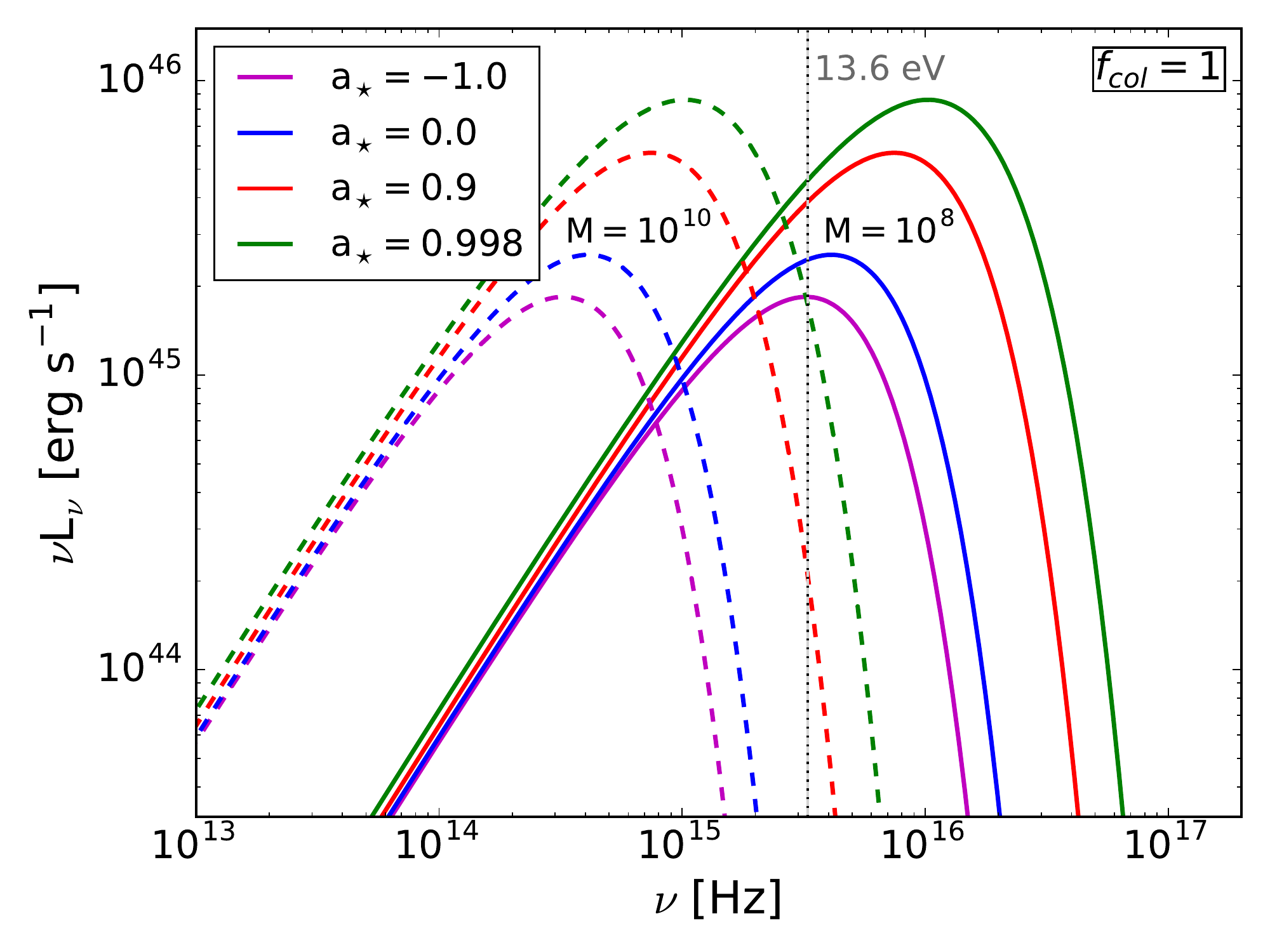}
\includegraphics[width=0.49\textwidth]{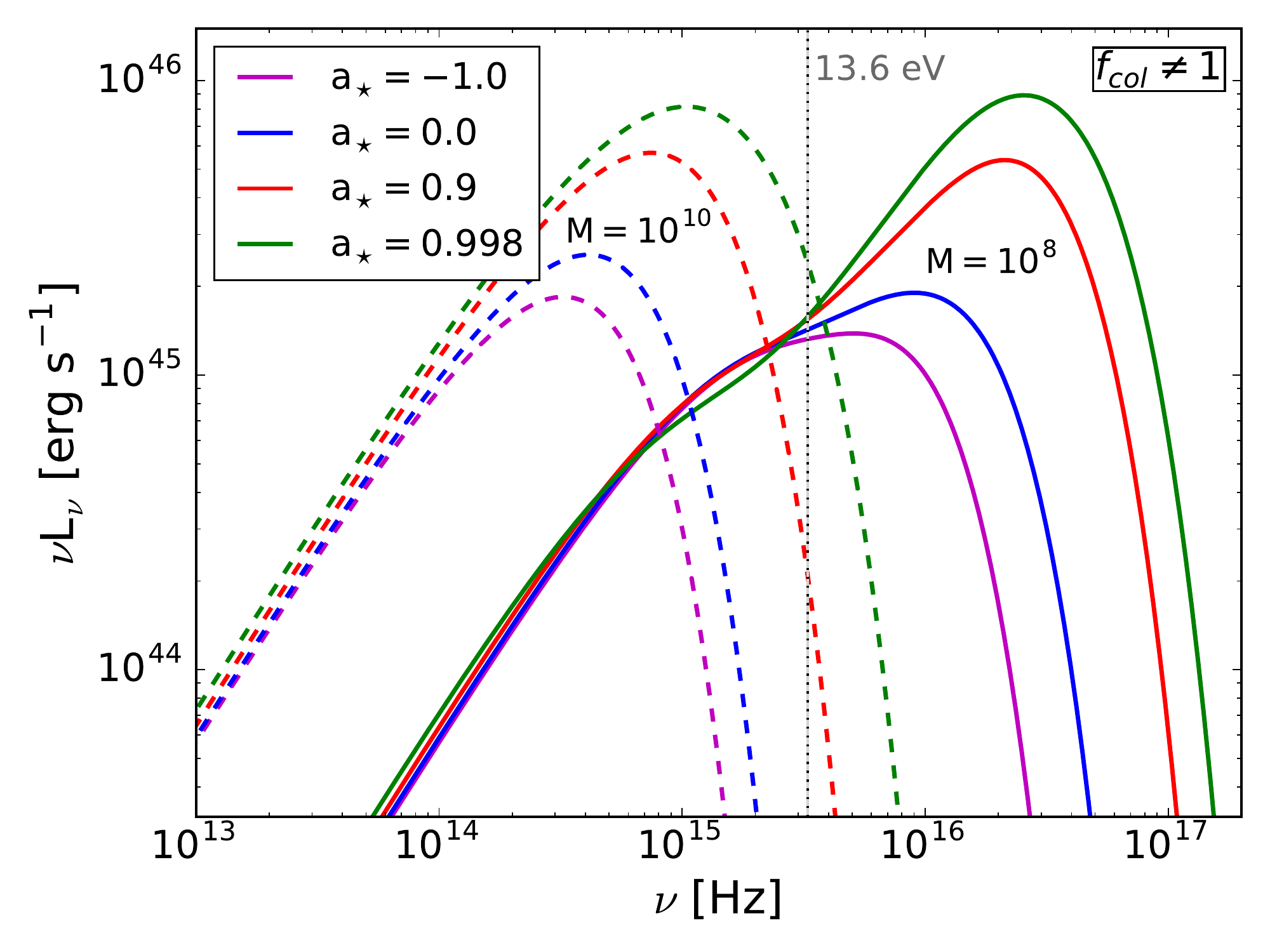}
\caption{Example model rest frame SEDs for different BH mass and spin values. 
The solid and dotted lines show models with $\mbh=10^{7}$ and $10^{9}\,\Msol$, respectively. All models assume $\dot{M}=1\,\mpyr$ and $\cos\left(inc\right)=0.8$. \textit{Left panel:} Simplified model SEDs that ignore temperature colour corrections (i.e., setting $f_{\rm col}=1$). \textit{Right panel:} Model SEDs including the self-consistent colour-temperature corrections provided by the \citep{Done2013} model.
}
\label{fig:DL_reproduced}
\end{figure*}

\begin{figure}
\centering
\includegraphics[width=0.49\textwidth]{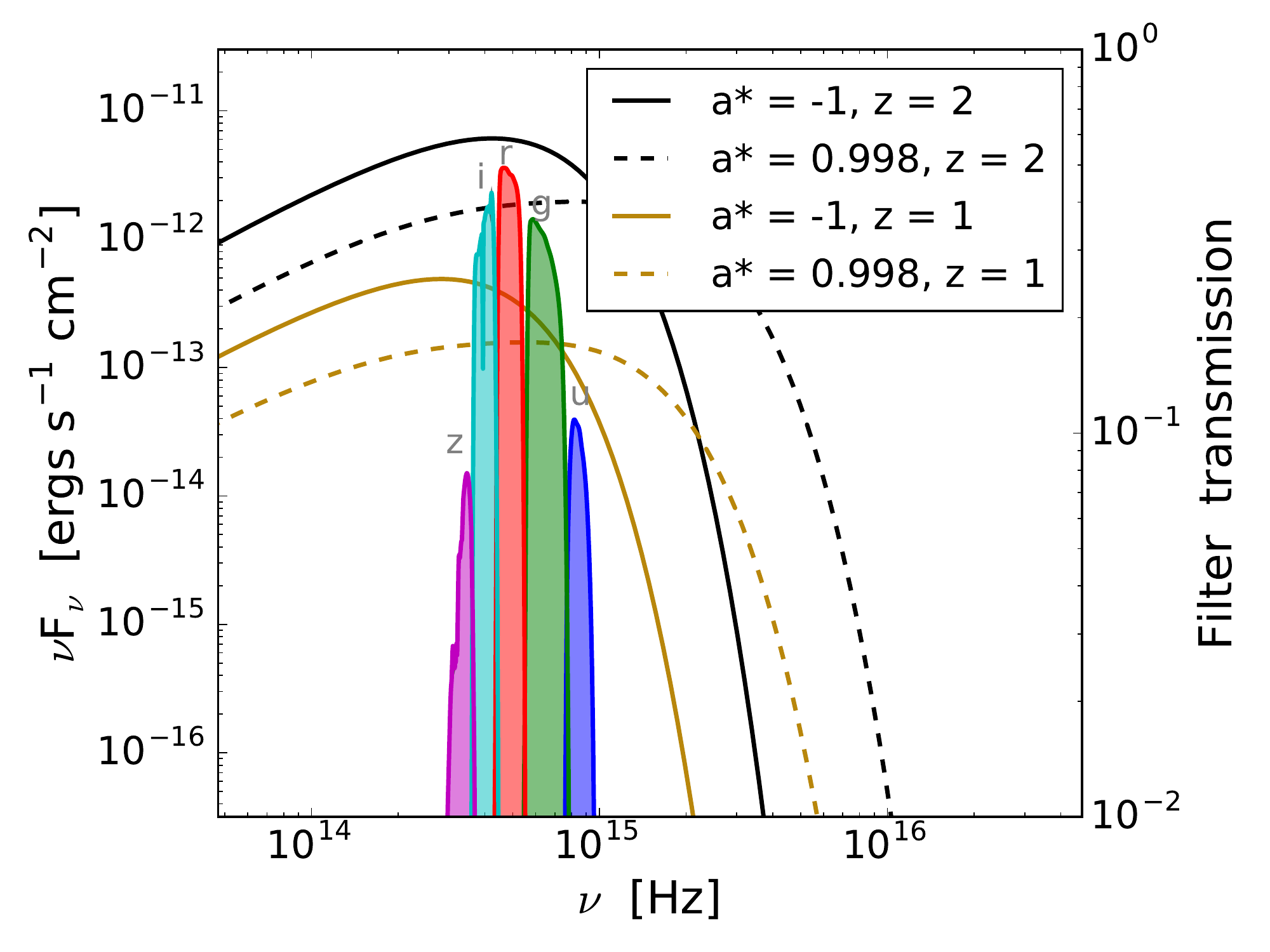}
\caption{Difference in the SED induced by changing \as from -1 to 0.998 and $z$ from 1 to 2 for a model with \mbh $= 10^{10}$ \Msol, $inc=10$ and \lledd\ $=0.1$ (so $\dot{M}=f(\lledd , \as, \mbh )$ is not constant and the SEDs don't overlap in their power-law, unlike in figure \ref{fig:DL_reproduced}). The SEDs are plotted in our observed frame. At this mass, $ugriz$ probes the turnover between power law and exponential cutoff more efficiently at higher redshift, and with that the effect of the spin. } 
\label{fig:filter_plot}
\end{figure}

\begin{table} 
\caption{Parameters used in the grid of model SEDs}
\begin{tabular}{l c c c}
\label{tab:Grid}
 & & & \\
\textbf{Parameter} & \textbf{Min. Value} & \textbf{Max. Value} & \textbf{Step Size} \\
\hline \\
BH mass, $\log\left(\mbh/\Msol\right)$ & $ 6 $ & $11$ & $0.5$ \\
 & & & \\
BH spin, \as  & $-1$ & $0.998$ & $0.1$ \\
 & & & \\
Accretion rate, \lledd  & $0.05 $ & $1$ & $0.05$ \\
 & & & \\
Redshift, $z$ & $0.5$ & $2$ & $0.1$ \\
 & & & \\
Inclination, $inc$ & $10\deg$ & $50\deg$ & $10\deg$ \\
 & & & \\
\end{tabular}
\end{table}

Our calculated model SEDs are based on the assumption of a geometrically thin, optically thick AD, emitting as a sum of blackbodies, originally developed by \citet{ShakuraSunyaev1973}, and have since been elaborated to include colour-temperature corrections and relativistic effects, among other improvements. 
Several recent studies demonstrate that thin AD models provide a good fit to the UV-optical continuum SEDs of unobscured AGN, particularly in cases where 
the SMBH is accreting at a fairly high fraction of the Eddington luminosity, and reliable estimates of \mbh\ are available  \cite[e.g.,][and references therein]{Jin2012a,Collinson2015_AD,Capellupo2015_ADs}.

{
We note that several quasar microlensing studies \cite[e.g.,][]{Pooley2007_microlensing,Dai2010_microlensing_AD,Blackburne2014_microlensing_AD}, as well as some evidence from reverberation mapping \cite[e.g.,][]{Fausnaugh2016_STORM_RM}, have challenged the expectations of the thin AD model, with the observations suggesting somewhat larger disc sizes. 
The present work, however, focuses on the UV-optical part of the continuum, here assumed to originate solely from the innermost disc regions, which most probably do extend towards the ISCO.
}

The model SEDs we use here were produced using the \texttt{optxconv} model presented in \citet{Done2013}, employed within the \texttt{xspec} package \cite[version 12.9.01][]{Arnaud1996_xspec}.  
Here we only mention the specific parameters we used for the purposes of the present work.
For a detailed discussion about the model, we refer the reader to \citet{Done2013}.
Our main grid of model SEDs is calculated using colour-temperature corrections, noted as $f_{\rm col}$, and affecting the intrinsic (blackbody) flux originating from each annulus in the disc following $f_{\nu}\left(T\right)=f_{\rm col}^{-4}\, B_{\nu}\left(f_{\rm col}\,T\right)$.
To verify that our results do not strongly depend on these corrections, we also consider an alternative grid of purely thin, Shakura-Sunyaev-like disc SEDs. 
In terms of the \texttt{optxconv} code parameters, the latter is achieved by setting $f_{\rm col}=1$.
Additionally, we don't consider coronal emission (setting $r_{\rm cor}=-1$), and we assume an arbitrarily large outer AD radius, $\log\left(r_{\rm out}\right)=7$. Both should not affect the UV-optical part of the calculated SEDs. 

We note that, unlike the public versions of  \texttt{optxagn} and \texttt{optxagnf}, the models we use here incorporate relativistic effects and an inclination dependence \cite[see][]{Done2013}.

We have created a grid of calculated accretion disc SEDs, spanning a wide range in several input physical parameters: BH mass \mbh, the BH spin \as, the Eddington ratio \lledd , the redshift $z$, and the inclination angle between the polar axis of the AD and the observer's line-of-sight $inc.$. 
The ranges and step sizes for all these parameters are given in Table~\ref{tab:Grid}. 
In total, the grid includes 369,600 unique model SEDs.
Our choices for the range in \mbh\ and \lledd\ are motivated by the results of several studies of large sample of quasars at $z\ltsim2$ \cite[e.g.,][]{Fine2008,Gavignaud2008_VVDS,McLure_Dunlop2004,Schulze2010_BHMF,Schulze2015_BHMF,Shen_dr7_cat_2011,TrakhtNetzer2012_Mg2,Trump2009b_MBH}. 
We have verified that our results do not depend strongly on our choice to use models with $\lledd>0.3$, for which the original \cite{ShakuraSunyaev1973} model is not expected to hold. 
Appendix~\ref{app:low_lledd} presents some of our main findings for a limited subset of model SEDs with $\lledd<0.3$.
The range in inclination angles is motivated by the requirement that the calculated SEDs would eventually be observed as type-I AGN.
We did not consider redshifts exceeding $z=2$, since the IGM-related absorption is expected to dominate the rest-frame UV part of the SEDs, observed within the SDSS imaging.
Figure~\ref{fig:DL_reproduced} presents the UV-optical part of several example model SEDs,    
with the left panel showing simplified thin disc SEDs ignoring colour-temperature corrections, and the right panel showing SEDs that inlcude self-consistent corrections. 
As can be seen, the corrections mainly affect SEDs of BHs with low masses and high spins. The simplified models are in excellent agreement with those presented in many other studies \cite[see, e.g., Fig.~1 in ][]{DavisLaor2011_AD}. 
{
As can be seen, the difference in UV emission between a maximally prograde spinning BH and a stationary BH is much larger than the difference between the maximally retrograde spinning case and the stationary case, since the change in ISCO is larger (see section \ref{sec:Intro}).
}
\\
Finally, we calculated synthetic magnitudes for each of the SEDs in the grid, in the five SDSS photometric bands (\textit{ugriz}), using standard procedures. 
Figure \ref{fig:filter_plot} displays the \textit{ugriz} filter curves alongside some of our SEDs.
We stress that, since each SED is calculated at a given redshift, the calculated synthetic magnitudes realistically reflect the \emph{continuum} flux level of the respective model quasar (i.e., ignoring contribution from emission lines), with no need for any arbitrary scaling.

Our fiducial model SEDs  do not take into account several additional effects that may alter the emergent SEDs and are expected in real AGN SEDs.
These include 
the hard X-ray emission and soft X-ray excess; 
outflows from the inner parts of the AD; 
dust extinction in the host galaxies; 
and additional emission from AGN-related lines and/or from the stars of the host galaxies.
These effects should be considered in light of our focus on the high (AGN) luminosity, high-\mbh\ regime, and particularly the sensitivity of the SDSS selection algorithm to BHs with low or retrograde spins.
We discuss each of these below.

The fraction of power dissipated in the hard X-ray corona and soft X-ray excess components should affect the inner disc spectrum if they are ultimately powered by the mass accretion flow. 
The energy required to power these X-ray emission components is then not available to power the standard disc emission, therefore making the disc continuum cooler and less luminous than predicted by a pure disc model \cite[e.g.,][]{Done2012}.
Our assumption of a pure disc SED is therefore a conservative choice, as it reflects no additional losses of UV radiation, and provides the best possible case for SDSS selection. 
Moreover, these additional X-ray components are expected to be weakest in the high-luminosity, high-\mbh\ regime which is our main focus in this work.

Dust extinction in luminous quasars is generally found to be limited (see, e.g., \citealt{Capellupo2016_XS_pap3}, \citealt{Baron2016_dust}, and references therein, but also the exceptional populations studied in, e.g., \citealt{Glikman2007} and \citealt{Banerji2012_UKIDSS}).
More importantly, dust extinction will reduce the UV emission from any calculated AD SED, in a way which broadly mimics the effects of reducing the BH spin parameter \cite[see, e.g., the discussion in][]{Capellupo2015_ADs,Collinson2015_AD}.
Outflows of gas from the inner parts of the AD would have a similar effect \cite[e.g.,][]{Slone2012}.
Thus, the inclusion of dust extinction and/or AD outflows may further limit the prospects of the SED being selected by the SDSS colour-based algorithm. 
Our choice to ignore these effects is, therefore, conservative in the context of the present work.

Both these may significantly reduce the radiation in the UV part of the SED, in a way which may broadly mimic the effects of reducing the BH spin parameter \cite[see, e.g., the discussion in][]{Slone2012,Capellupo2015_ADs,Collinson2015_AD}.
These may further complicate the emergent UV SED, but require a set of additional assumptions and parameter choices (e.g., outflow profile, extinction curve), which are beyond the scope of the present work.

Emission lines and features would affect the colour selection only if they are strong (i.e., in terms of equivalent width), and contribute differentially to two SDSS bands (or more). 
We have verified that our analysis is not significantly affected by the inclusion of emission lines, following the typical emission-line spectrum seen in SDSS quasars. 
We describe this test and the associated results in Appendix~\ref{app:em_lines_test}.
We also note that in the high-\mbh, low- (or retrograde-) spin regime, the UV emission lines are actually expected to be \emph{weaker} than usual, due to the low number of ionizing photons, thus further limiting their possible effect on colour-based selection.

Finally, for high-luminosity SDSS quasars at $z\gtrsim1$ the host galaxy contribution is expected to be low, of order $\ltsim25\%$ at rest-frame 3000 \AA\ \cite[][]{Schulze2015_BHMF}, and significantly lower at yet shorter wavelengths \cite[e.g.,][]{Merloni2010,Collinson2015_AD} -the spectral regime where the effects of \as\ are expected to be most dramatic.

\subsection{Overview of the SDSS Quasar Selection criteria}
\label{sec:SDSSCriteria}

The original SDSS project covered about $10,000$ deg$^{2}$ in its main imaging survey, as well as a smaller area of $\sim$750 deg$^{2}$, in five different bands \citep{Doi2010}, \textit{ugriz}.
Quasars are selected for (multiplexed) follow-up spectroscopy through an elaborate algorithm, discussed in detail in \citep{Richards2002}.
The algorithm is designed to provide both high spectroscopic completeness \emph{and} purity for quasars, building on the rich experience gained from previous optical spectroscopic quasar surveys, and incorporating the commissioning data of the SDSS itself. 
However, to the best of our knowledge, the algorithm was never tested against physically-motivated models of quasar continuum emission.

The quasar selection algorithm is primarily based on selection in a 4-dimensional colour-colour space, defined by $u-g$, $g-r$, $r-i$, and $i-z$ colours. 
The 4-dimensional colour space is further divided into two different 3-dimensional colour-colour selection sub-spaces: 
the $ugri$ bands are used for identifying low-redshift quasar candidates, the $griz$ bands are probing at longer wavelengths and thus more suitable for higher redshifts.
A separate selection procedure focuses on radio-emitting sources, based on a cross-match to the large-area FIRST survey \cite[][]{Becker1995}, but provides only a small minority of all  quasar candidates which are not also otherwise colour-selected \cite[336 out of 8630; e.g.,][]{Schneider2010_QSOCAT_DR7}.
Some (four-dimensional) regions in the colour-colour space are especially prone to contamination by white dwarfs, A stars, and M star - white dwarf binaries. 
The latter are defined as exclusion regions and any objects within them are immediately rejected (see red regions in Figure \ref{fig:cc_plots}).  
On the other hand, any objects lying within the defined inclusion regions (green regions in Figure~\ref{fig:cc_plots}) are directly selected, except if they were already rejected by the exclusion boxes, which have a higher priority. 
We emphasise that there exists one unique inclusion box: the UV-excess (``UVX'') selection region is indeed a two-dimensional region and not the projection of a four-dimensional region.
The region of the colour space that is generally occupied by stars (and normal galaxies) -- the so-called called ``stellar locus'' (excluded from Figure~\ref{fig:cc_plots} for clarity) -- defines another significant (multi-dimensional) exclusion region. 
The stellar locus was modelled explicitly by \cite{Richards2002} and kept constant during the survey. 

Any objects that lie sufficiently far outside the stellar locus, in either the \textit{ugri} or the \textit{griz} projections, and do \emph{not} fall within any of the aforementioned exclusion regions, are also selected as targets, even if are not located in any specific inclusion region. 

The selected objects will undergo several quality tests and, finally, if they are within the pre-defined optical flux limits, they will be marked as possible quasar candidates. 
In order to be included within the SDSS flux limits for follow-up spectroscopy, the object must have an $i$-magnitude between 15 and 19.1, for the low-redshift ($ugri$) selection criteria, or between 15 and 20.2 for the high-redshift ($griz$) ones.
We note that the latter criteria may, occasionally, select low-redshift sources as well.

\section{Results and Discussion}
\label{sec:ResultsDiscussion}

We now turn to analyse the location of the model SEDs in the SDSS colour-colour space, with the goal of quantifying their prospects for selection for spectroscopy by the colour-based target selection algorithm. 
We particularly focus on the role of \as\ in the high-\mbh\ regime.

We note that testing the relevance of thin AD SEDs for \emph{real} AGNs (observed in SDSS and other surveys), for which some physical parameters are determined, is beyond the scope of the present work.
This subject has been addressed in several recent works, with somewhat ambiguous results.
For example, the recent studies by \cite{Capellupo2015_ADs,Capellupo2016_XS_pap3} and \cite{Collinson2015_AD} showed good agreement between the UV-optical spectra of luminous AGNs, which have reliable estimates of \mbh, and the corresponding thin AD SEDs. 
The study of \cite{Jin2012a} successfully extended this approach to account for the X-ray emission.
On the other hand, \cite{Davis2007_UVcont} found some discrepancies between the observed UV SED shapes and what is expected from thin AD models.

\subsection{Model SEDs in the colour-colour space}

\begin{figure*}
\centering
\setlength{\fboxsep}{8pt}
\fbox{
\stackon{
\includegraphics[width=0.32\textwidth]{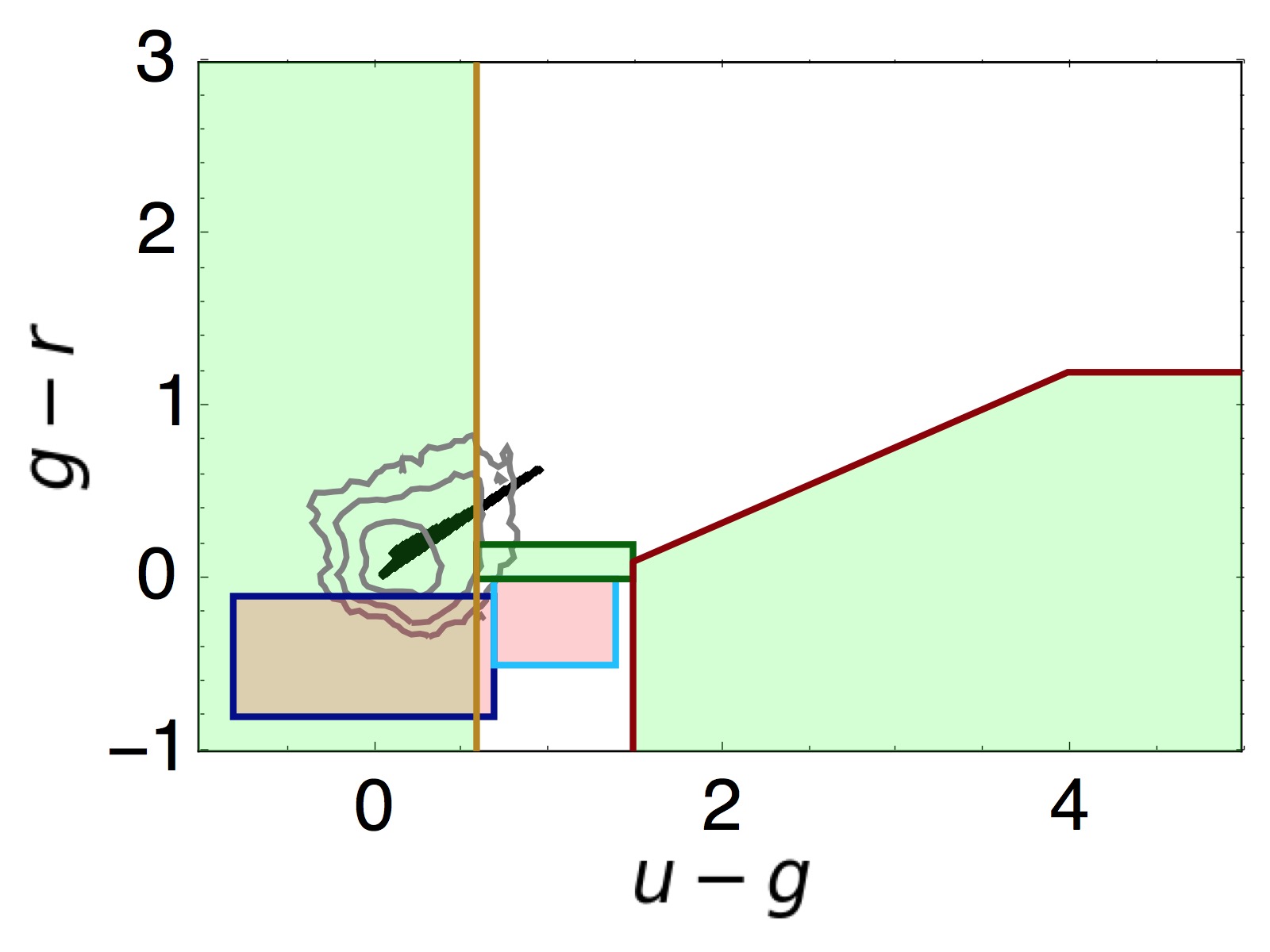}
\includegraphics[width=0.32\textwidth]{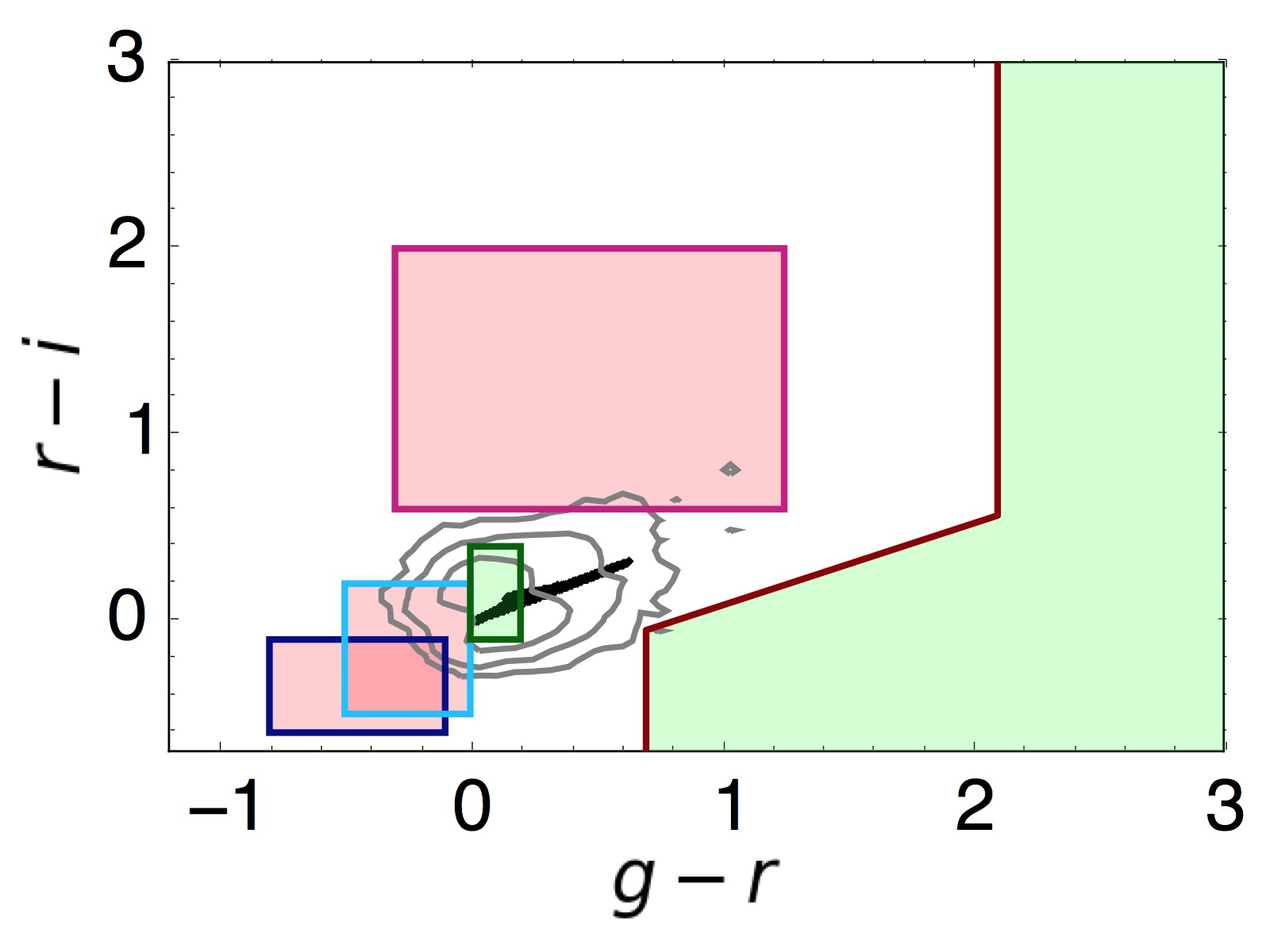}
\includegraphics[width=0.32\textwidth]{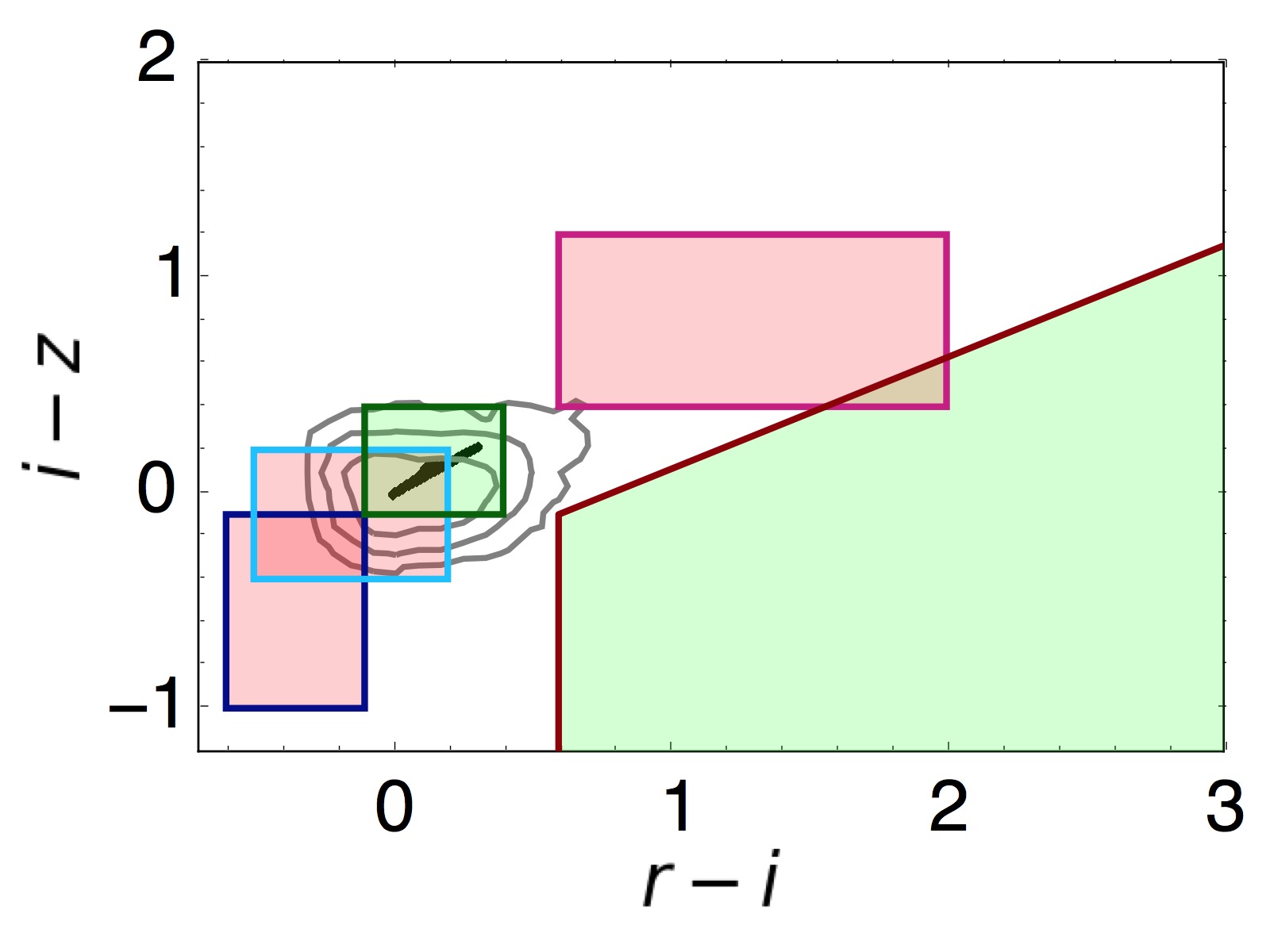}
}{a) Models ($15 \leqslant i \leqslant 20.2$, \mbh $\leqslant 10^{9.5}$ \Msun) vs. colour-selected DR7 quasars}
}
\par
\vspace{0.5cm}
\par
\fbox{
\stackon{
\includegraphics[height=0.25\textwidth]{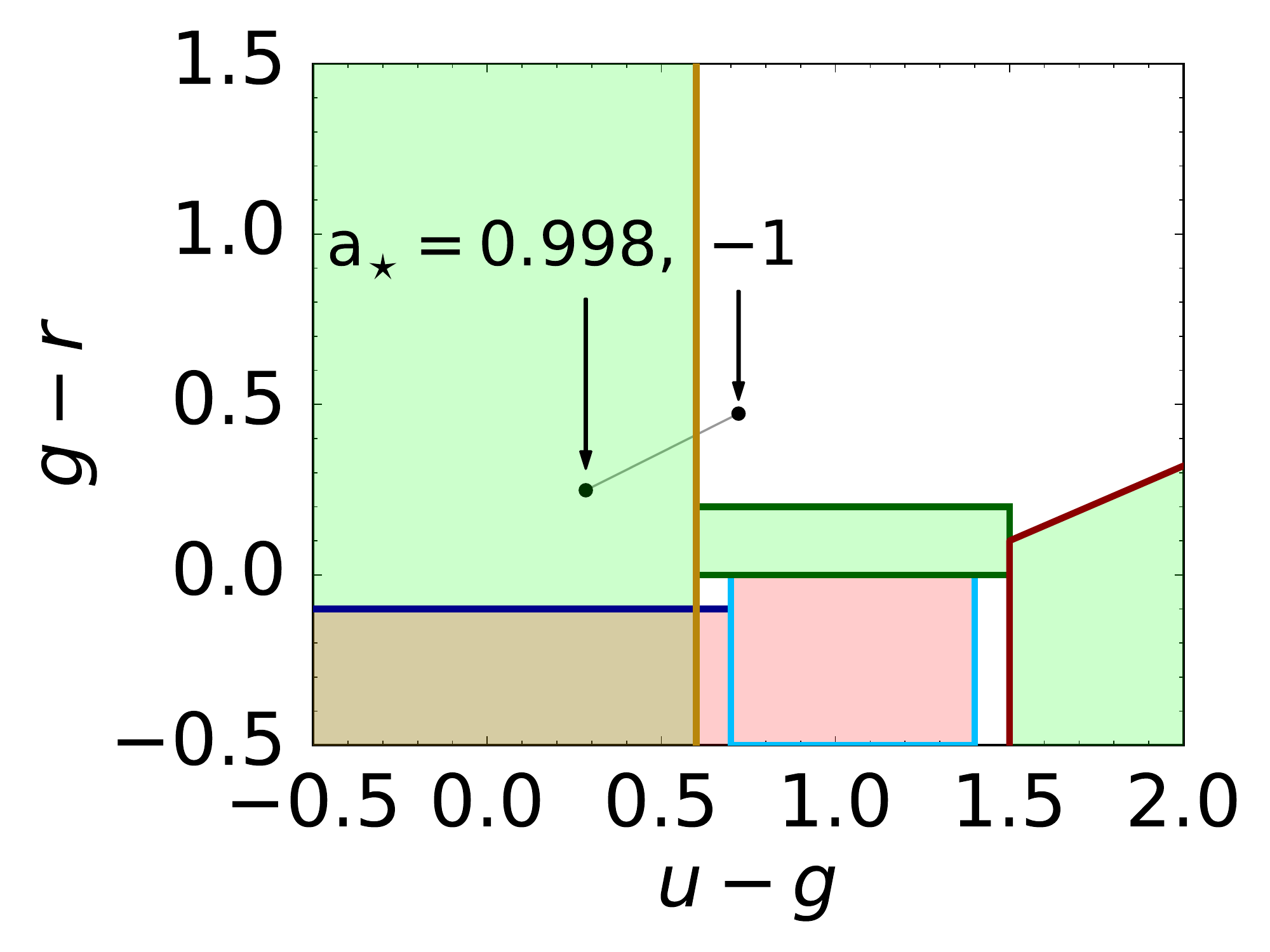}
\includegraphics[height=0.24\textwidth]{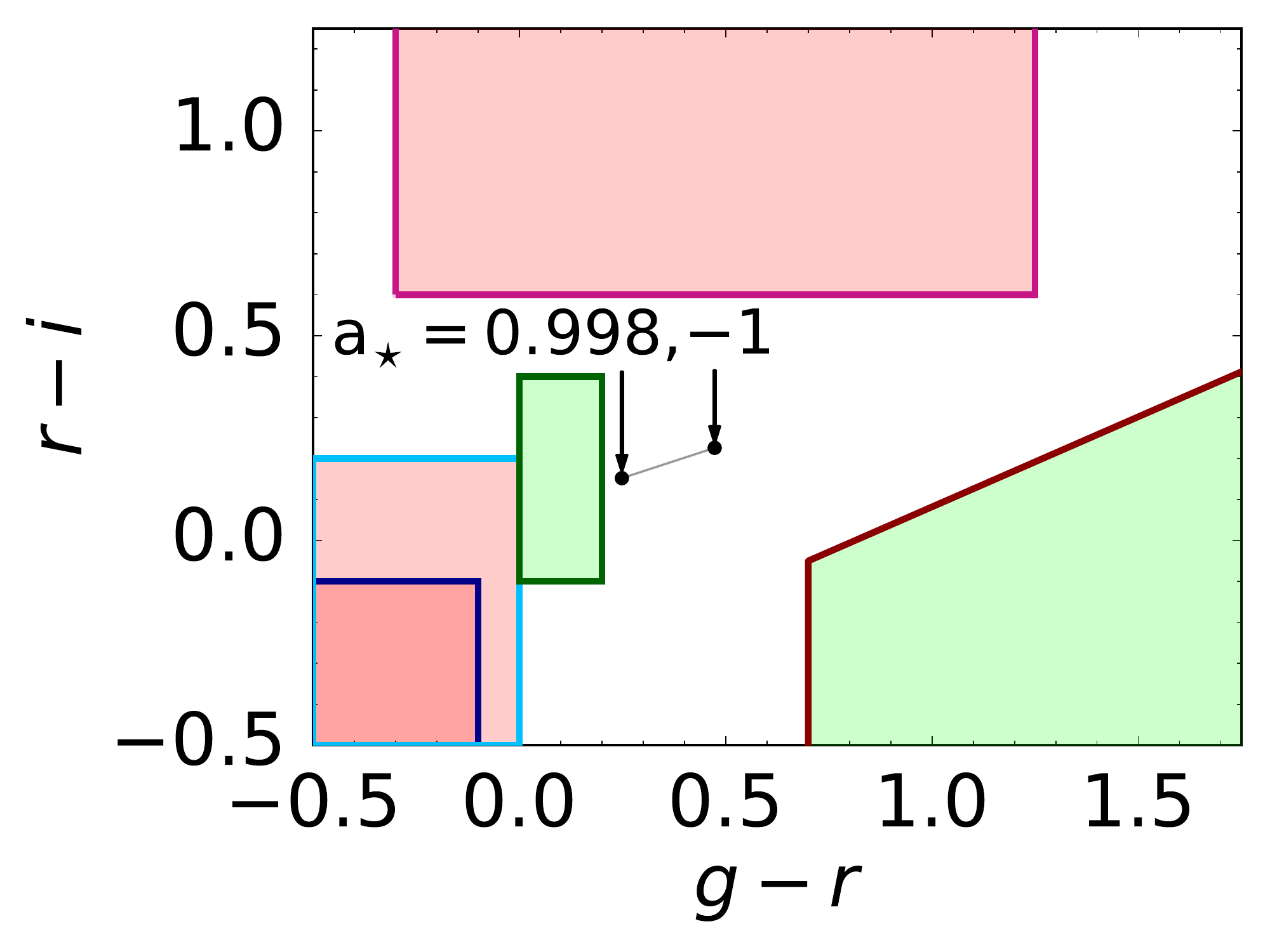}
\includegraphics[height=0.24\textwidth]{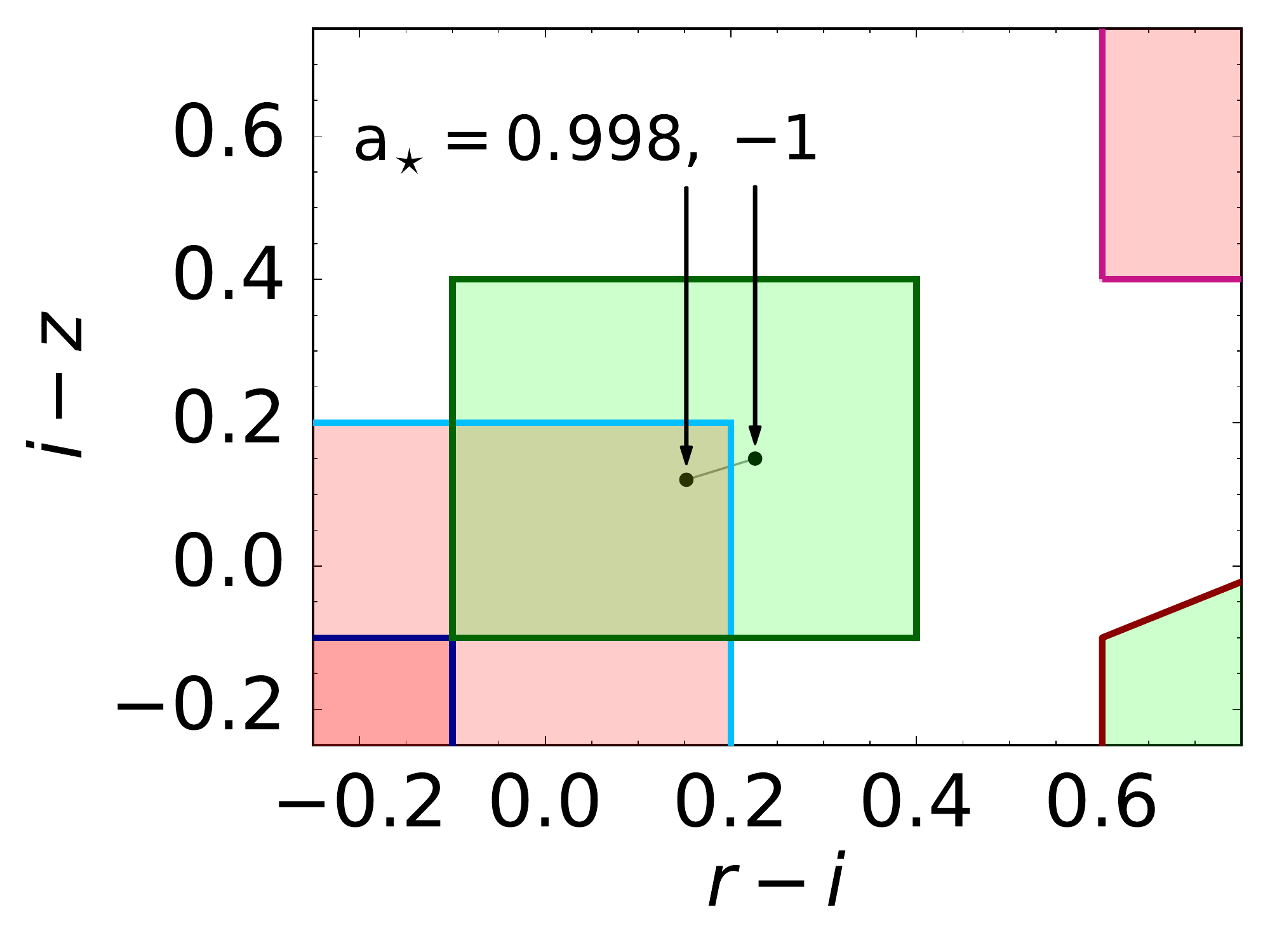}
}{b) Effect of changing BH spin, for a given AGN ($\mbh = 10^{10} \Msun$, $z=1$, $\lledd=0.1$, $inc=10^{\circ}$)} }
\par
\vspace{0.5cm}
\par
\fbox{
\stackon{
\includegraphics[height=0.25\textwidth]{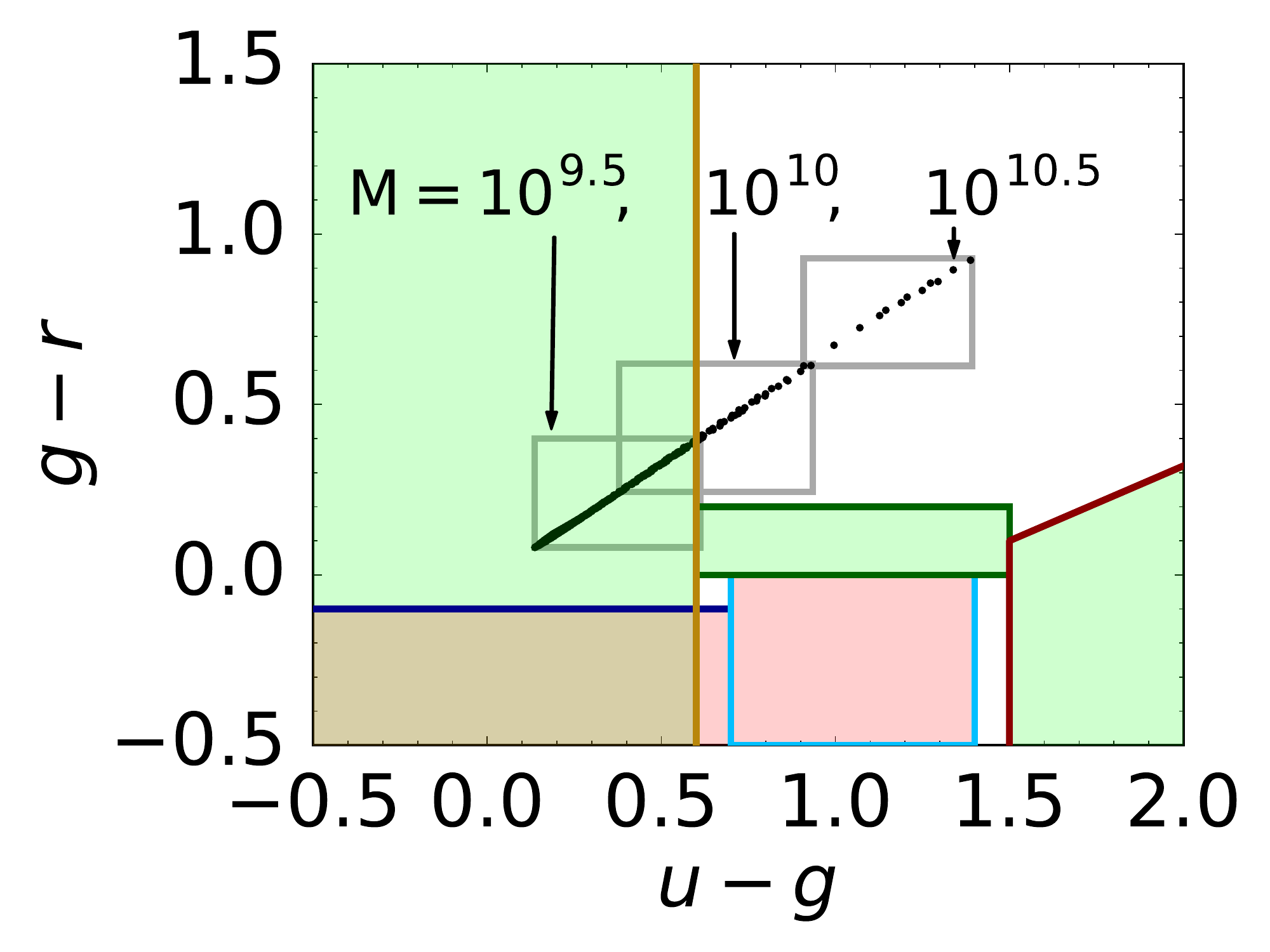}
\includegraphics[height=0.24\textwidth]{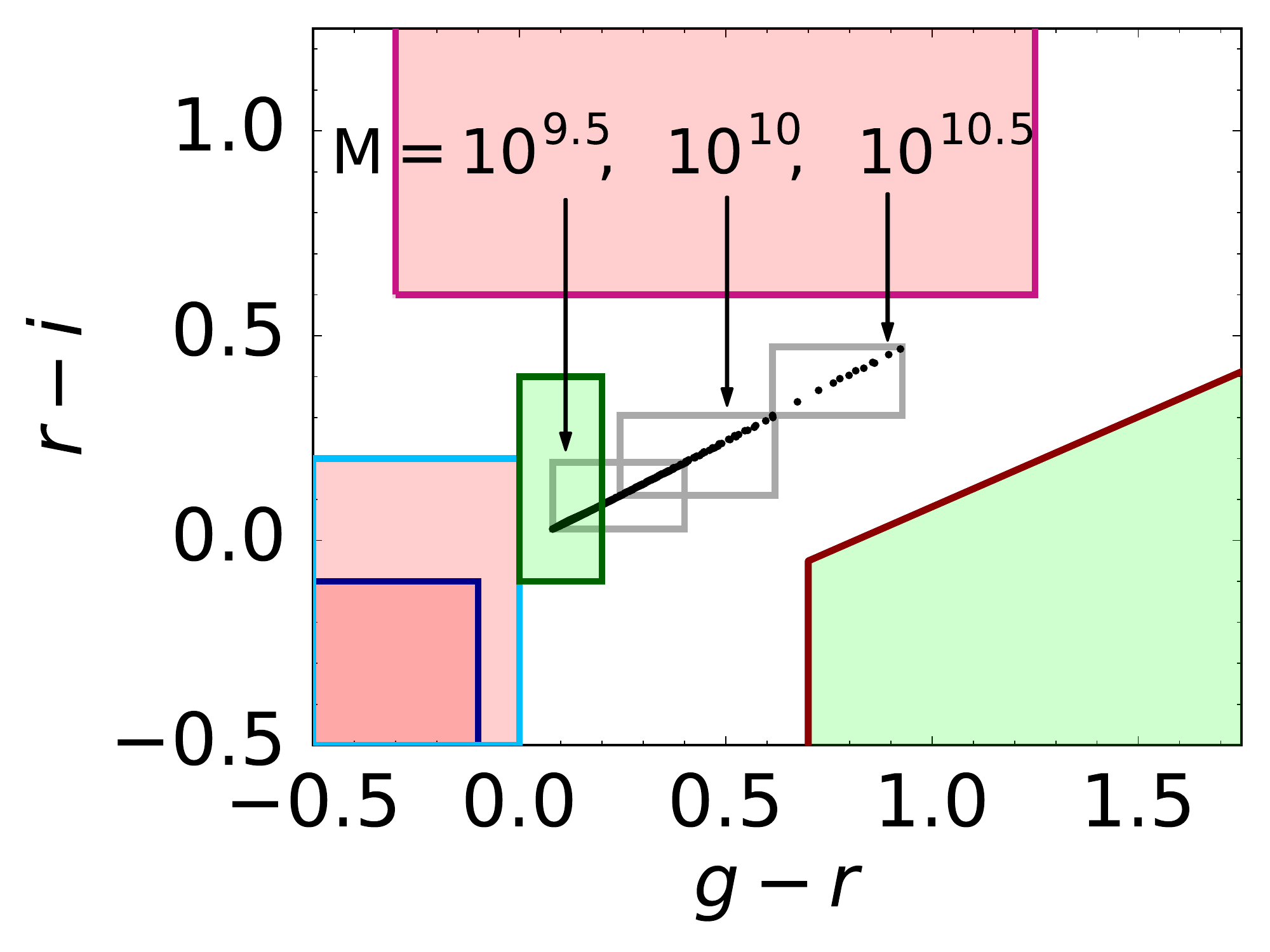}
\includegraphics[height=0.24\textwidth]{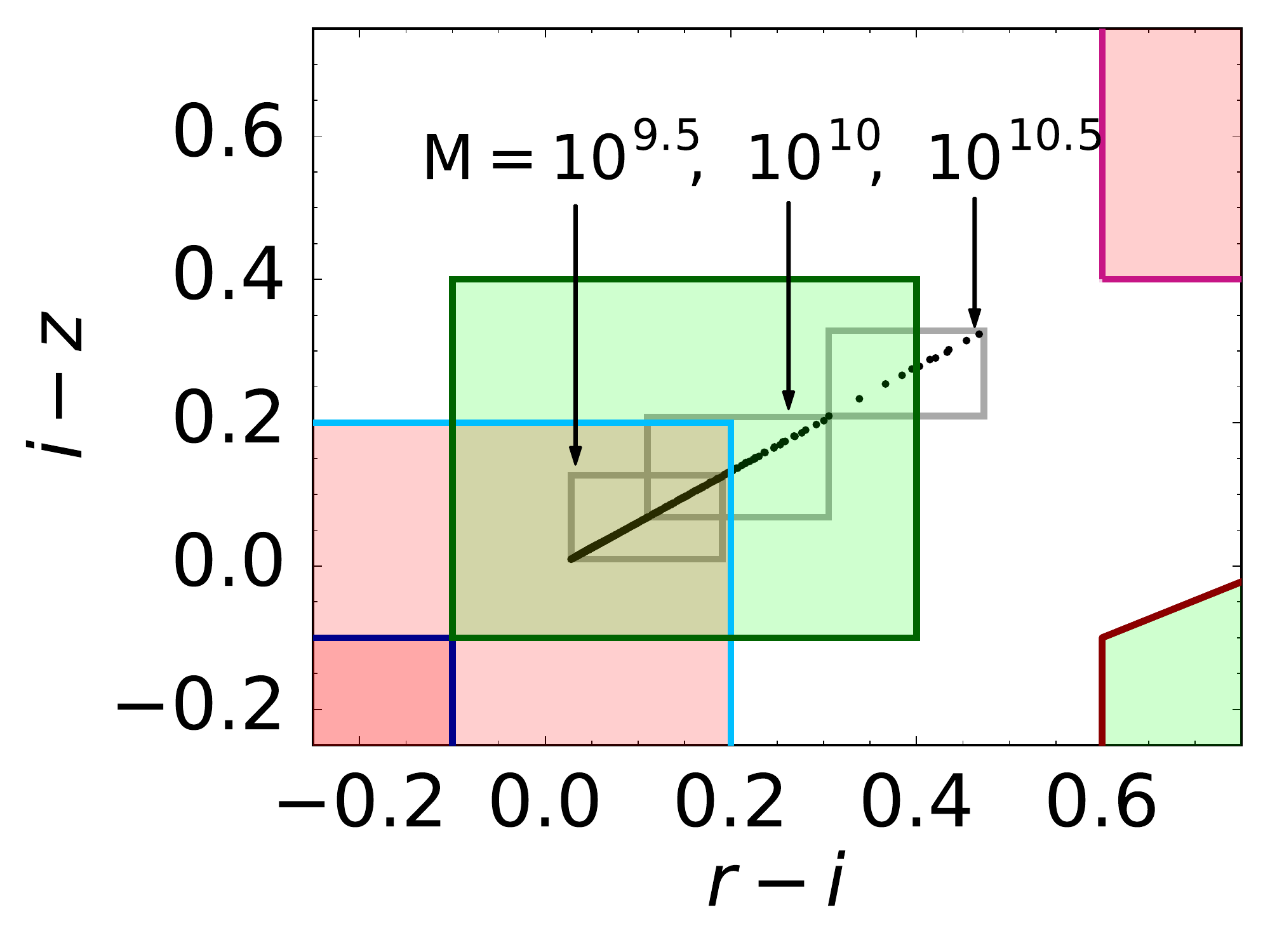}
}{c) Effect of increasing BH mass, for a population of AGN ($\as = -1$, $0.5 \leqslant z \leqslant 1$, all \lledd , all $inc$) } 
}
\par
\vspace{0.5cm} 
\par
\caption{Position of our model SEDs in the SDSS colour-colour space. 
In all panels, green and red shaded regions mark the inclusion and exclusion regions, respectively, of the SDSS colour-based quasar selection algorithm. 
All of the model SEDs shown here include colour-temperature corrections. Note that the panels in the middle and bottom rows show an insert of the full colour-colour space (i.e., of the top row panels).
\textit{Top three panels}: 
blue points trace all the models that are within the SDSS flux limits ($15 \leqslant i \leqslant 20.2$), and with $\mbh \leqslant 10^{9.5}\, \Msun$. 
The contours represent the colour-selected, $0.5 \leqslant z \leqslant 2$ quasars actually observed and classified in the SDSS/DR7 quasar catalogue, with subsequent contours tracing order-of-magnitude steps in the number of encircled objects.
\textit{Middle three panels}: the change in the position in colour-colour space caused by varying \as\ from -1 to 0.998, for a model with $\mbh = 10^{10}\,\Msol$, $\lledd=0.1$ and $inc.=10\deg$, at redshift $z=1$. 
In the left panel, the model with $\as=0.998$ lies within the UV-excess inclusion region while the model with $\as=-1$ drops out of this region, and is therefore not selected for spectroscopy.
\textit{Bottom three panels}: the change in the position in colour-colour space caused by varying the BH mass, while keeping BH spin fixed at $\as=-1$. 
The grey rectangles bracket models with $\mbh=10^{9.5}$, $10^{10}$, and $10^{10.5}\,\Msun$ (as illustrated), at redshifts $0.5 < z < 1$. 
}
\label{fig:cc_plots}
\end{figure*} 

In this section, we discuss the position of our model SEDs in colour-colour space and with respect to the relevant inclusion and exclusion regions.
We focus on specific examples where changing the BH spin parameter has a significant effect on the outcome of the selection algorithm.
In what follows, we mainly focus on the set of model SEDs that include colour temperature corrections.

The top three panels in Figure~\ref{fig:cc_plots} (a) present the location of a large subset of our model SEDs in the SDSS colour-colour space.
In these panels we only show models fulfilling $15 \leqslant i \leqslant 20.2$ and $\mbh \leqslant 10^{9.5}\, \Msun$ (as blue points). 
We have not included the models with larger masses in order to allow for a better comparison to actual observations. 
We also show the observed population of colour-selected quasars with $0.5 \leqslant z \leqslant 2$ from the SDSS/DR7 quasar catalogue \cite[coloured contours][]{Abazajian2009_DR7, Schneider2010_QSOCAT_DR7}. 
For our models, we adopt the deeper flux limit of $15 \leqslant i \leqslant 20.2$ defined for the SDSS \textit{griz}-selection, while most of SDSS sources at $0.5 \leqslant z \leqslant 2$ (about 90\%) are actually selected via the \textit{ugri}-based selection criteria, and therefore subjected to the more limiting condition $15 \leqslant i \leqslant 19.1$. 

As can be seen in the top panels of Fig.~\ref{fig:cc_plots}, our models generally lie within the space occupied by the observed SDSS quasars.
Moreover, $\sim 99.6$ \% of them are located within the UV-excess (``UVX'') inclusion region ($u-g<0.6$), and none of them are found within any (multi-dimensional) rejection region.

However, the colour-selected SDSS/DR7 quasars occupy a somewhat more extended region in colour-colour space, 
particularly in the direction perpendicular to the sequence formed by our models. 
The fact that our models do not account the whole span of real quasars is probably due to the varying degree of contamination from (mainly broad) emission lines, and/or the quasars' host galaxies \cite[see, e.g.,][]{Hao2013}.

The centre row of panels in Figure \ref{fig:cc_plots} (b) illustrate the impact of the BH spin on the location of SEDs in SDSS colour space. 
This is illustrated by changing only the spin parameter, between $\as = -1-\, \left(+\right)0.998$, for a subset of models where the other parameters are fixed at $\mbh = 10^{10}\,\Msol$, $\lledd=0.1$, and $inc.=10\deg$.
As can be seen, the change in \as\ shifts the position of the model SEDs in the colour-colour plots, and in certain cases may make the difference between the model SED being selected or not, as we further discuss in Section~\ref{sec:Completeness}. 
In this example, the model with \as $=0.998$ would be selected through the UV excess inclusion region (green region with golden border in the left panel), while the model with \as $=-1$ is not selected since it is not located within \emph{any} of the inclusion regions, and is not positioned sufficiently far outside the stellar locus.\footnotemark
\footnotetext{We note that none of the models in this example are rejected due to their position in the light blue rejection box in the right panel, since the latter constitutes the projection of a four-dimensional region. In order to be rejected, an object would have to lie within the regions with a light blue frame in the other two panels as well.} 
The bottom three panels of Figure~\ref{fig:cc_plots} (c) focus on the effect of increasing mass in the high-mass regime, which will be discussed further in the next section.

\subsection{Completeness}
\label{sec:Completeness}

\begin{figure*}
\centering
\includegraphics[height=0.41\textwidth]{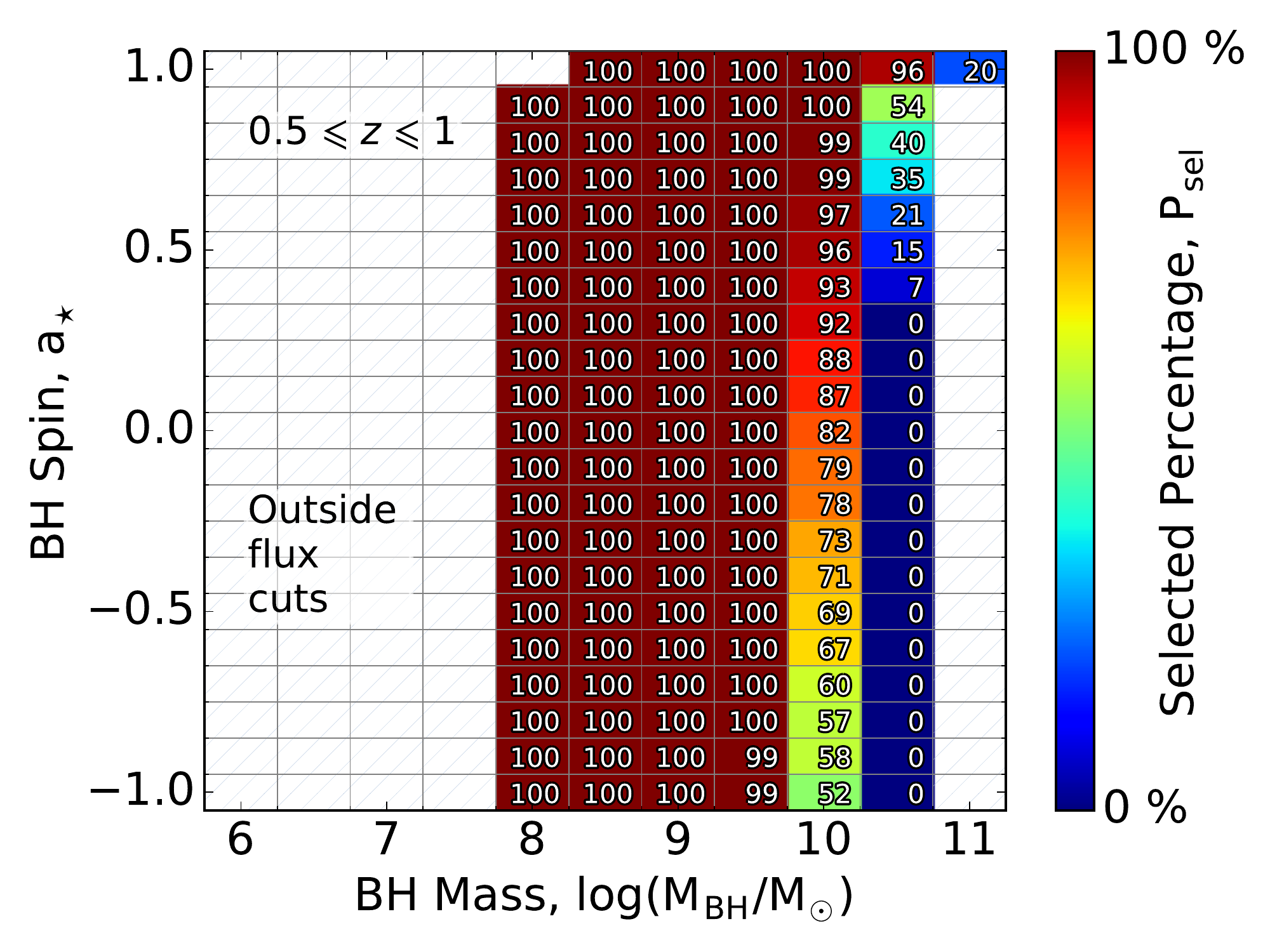}
\includegraphics[trim={0 0 4.1cm 0},clip,height=0.41\textwidth]{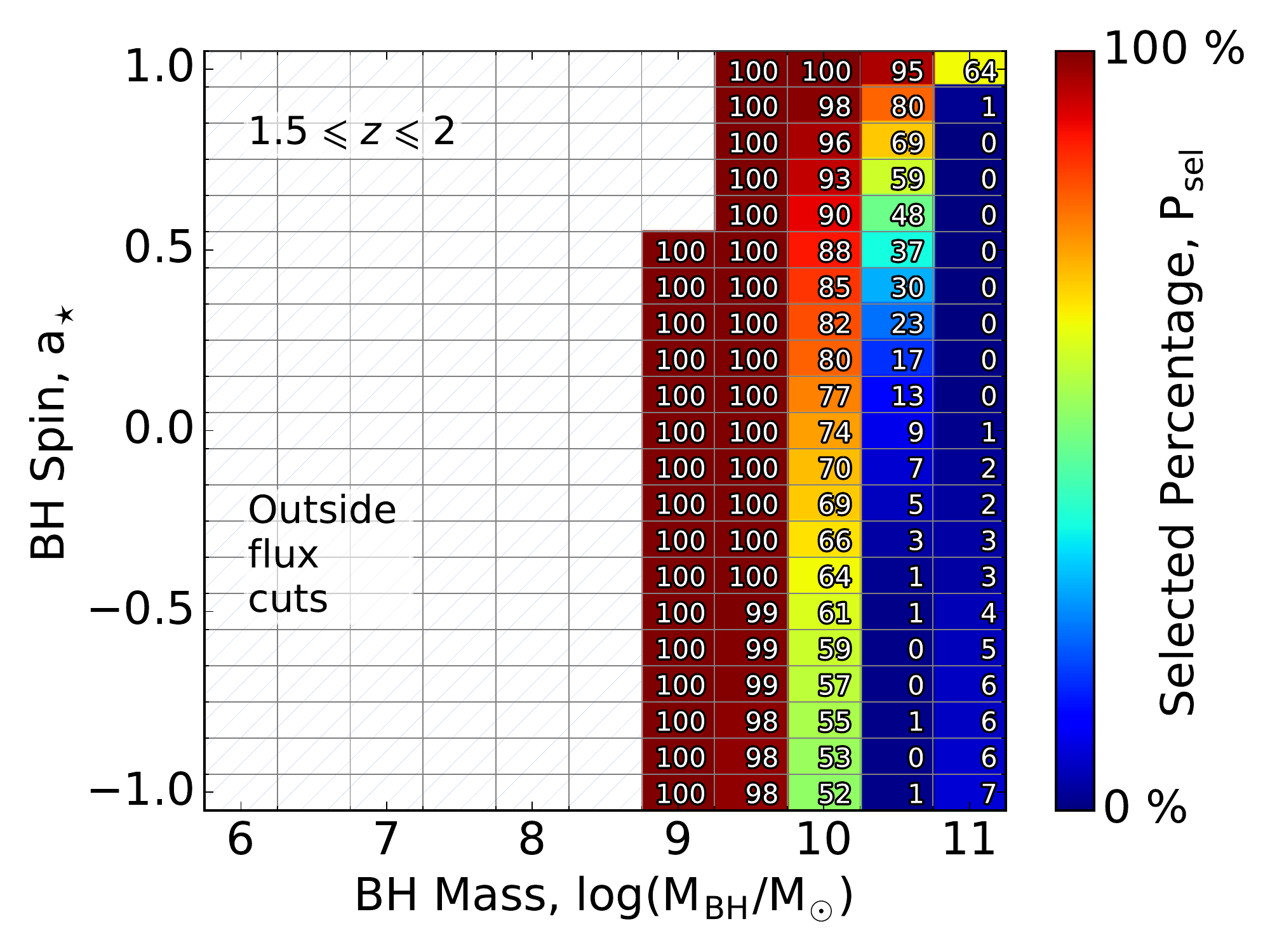} 
\caption{Percentage of \emph{observable} sources (i.e., within the flux limits) that are selected as possible quasar candidates in each bin of $\left(\mbh,\as\right)$. 
Hatched bins indicate that all of the objects lie outside SDSS's flux limits.
We stress that the number of observable objects varies between adjacent bins, even if they result in identical percentages of colour-selected objects (see Section~\ref{sec:Completeness}).
}
\label{fig:Percentages1}
\end{figure*} 

\begin{figure*}
\centering
\includegraphics[height=0.41\textwidth]{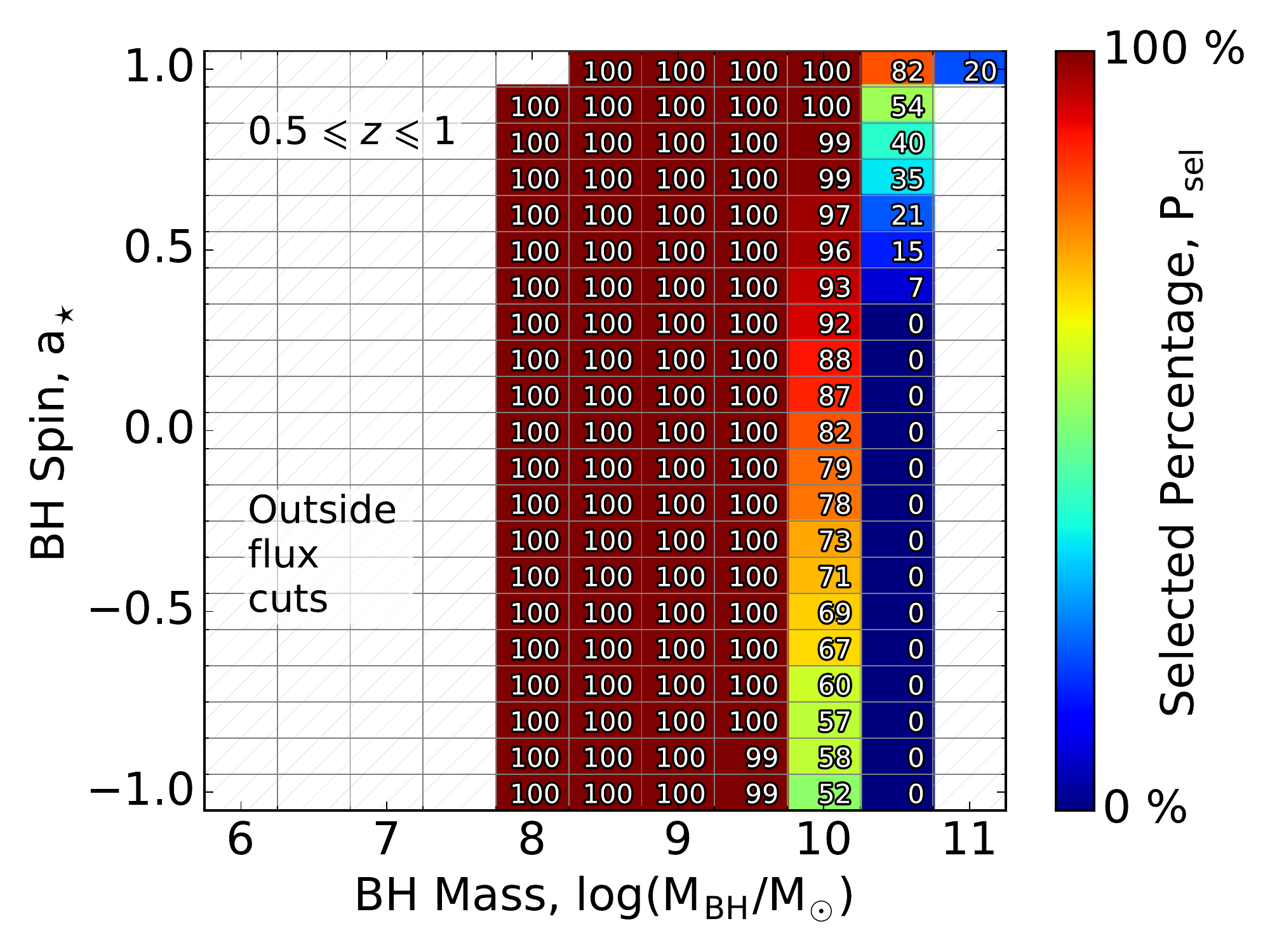}
\includegraphics[trim={0 0 4.1cm 0},clip,height=0.41\textwidth]{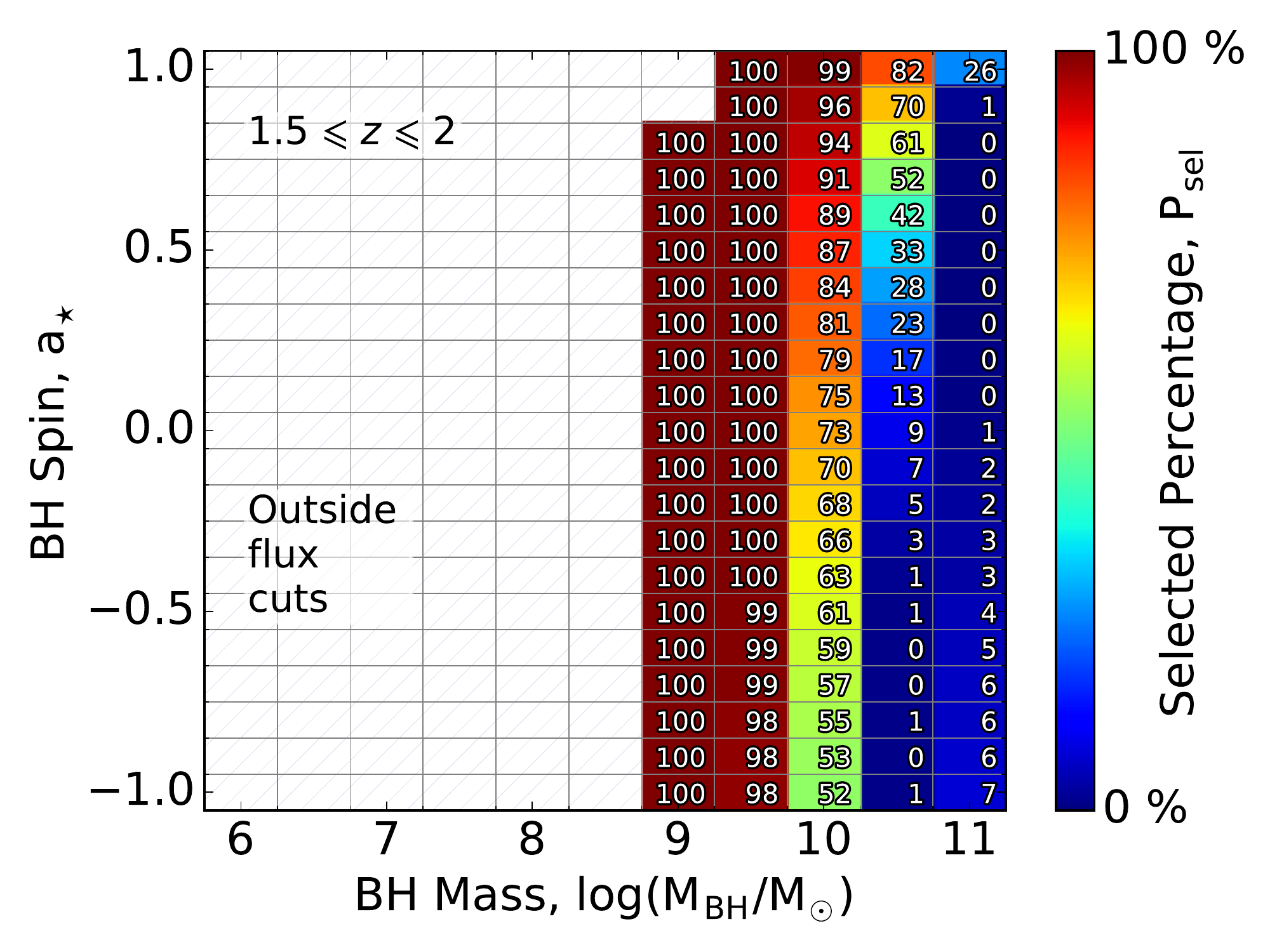} 
\caption{Same as Figure~\ref{fig:Percentages1}, but for simplified model SEDs that ignore temperature colour corrections (i.e., setting $f_{\rm col}=1$ in the \citep{Done2012} model).}
\label{fig:Percentages1_B}
\end{figure*}

We next examine what fraction of our model SEDs are selected as possible quasar candidates by the SDSS colour selection algorithm, across the range in SMBH properties. 
We focus only on ``observable'' model SEDs, defined through the relevant SDSS spectroscopic flux limits.
We define a model as being \emph{unobservable} either if it fails to satisfy the flux limits of the \textit{griz}-channel, [15, 20.2], or if it is not chosen by any of the \textit{griz}-channel criteria (so can only be selected by \textit{ugri}-space tests), but is also not included in the shallower \textit{ugri} flux limits, [15, 19.1] (see Section~\ref{sec:SDSSCriteria}).

We first find that for model SEDs with $\mbh \leqslant 10^{9.5}\, \Msun$, the selection algorithm is extremely efficient, with a completeness rate of 99.8\% (when including all values of  \as, \lledd, $inc$ and $z$, as listed in table \ref{tab:Grid})
All of the latter are selected by their UV excess through SDSS's UVX test, and about $0.014$ \% are additionally selected by SDSS's ugri test. 

Next, we consider each bin of a given BH mass and spin, individually, to find out what percentage of the observable models within it are selected. 
Each such bin contains the observable model SEDs from a three-dimensional grid ranging over different values of \lledd , $inc$ and one of two different redshift ranges - $0.5 \leqslant z \leqslant 1$ and $1.5 \leqslant z \leqslant 2$. 
Figure \ref{fig:Percentages1} shows the resulting completeness rates for all the fiducial model SEDs, while Figure~\ref{fig:Percentages1_B} focuses on models ignoring colour-temperature corrections. 
We stress that the percentages shown in Figs.~\ref{fig:Percentages1} and Figure~\ref{fig:Percentages1_B} (and also \ref{fig:Percentages2} and \ref{fig:Percentages2_B} in Appendix~\ref{app:low_lledd}; see below) trace the fraction of model SEDs that would have been selected by the SDSS colour-selection algorithm among all \emph{observable} objects (i.e., those that satisfy the general $i$-band flux limits). This means that (\mbh,\as) bins with identical percentages generally vary in the number of observable, and thus selected model SEDs. 
This is of particular importance for the lowest-mass, observable BHs ($\mbh=10^{8}$ and $10^{9}\,\Msol$, for the lower- and higher-redshift ranges, respectively), where the number of observable model SEDs drops monotonically with increasing \as, but the percentage of those SEDs that would be colour-selected remains 100\%.\footnote{The decrease in the number of observable objects with an increasing \as, for a fixed \mbh\ and (range of) \lledd, is due to SED becoming harder, while maintaining a fixed (range of) \Lbol. This corresponds to lower optical luminosities and therefore fainter $i$-band fluxes. See Figure~\ref{fig:filter_plot} for an example (c.f. the solid and dashed lines).}

As can be clearly seen in Figure~\ref{fig:Percentages1} and \ref{fig:Percentages1_B}, at the high-mass regime, where $\mbh \geqslant 10^{10} \Msun$, the percentage of selected SEDs (i.e., the completeness) decreases significantly with mass and, more importantly, toward low or retrograde (negative) BH spin parameters, at fixed BH mass.
This general trend persists both at our lower and higher redshift ranges, as it does in models with or without colour-temperature corrections, and and in the subset of models with $\lledd \leqslant 0.3$ (as shown in Figures \ref{fig:Percentages2} and \ref{fig:Percentages2_B} in the Appendix).
For example, for $\mbh = 10^{10}\, \Msun$, the fraction of selected observable models at $z\simeq0.8$ (i.e., left panels of Figs~\ref{fig:Percentages1} and \ref{fig:Percentages2}) drops from $\sim$100\% for $\as=0.998$, to $\sim$80\% and eventually $\sim$50\%, for models with $\as\simeq 0$ and $-1$, respectively. 
For SEDs with lower \lledd\ at higher redshifts (i.e., right panel in Fig~\ref{fig:Percentages2}), the fraction drops more dramatically, reaching $\sim$20\% for $\as\simeq0$ and single-digit percentages at $\as\leqslant-0.5$. 
At this high mass, a maximally retrograde spinning active BH would thus be only about half as likely to be selected as a candidate than a maximally prograde spinning one (at most).
Such a trend in the extremely high-mass regime is relevant because this is where probing the spin could allow us to learn something about the BH's accretion history (see section \ref{sec:Intro}).

We will now get back to the position in colour-colour space for some of the most relevant models to see what causes them to fail to be selected.
The bottom panels in Figure~\ref{fig:cc_plots} (c) focus on lower-redshift model SEDs which have a fixed, maximum-retrograde spin parameter $\as=-1$ and masses of $\mbh=10^{9.5}$, $10^{10}$ and $10^{10.5}\,\Msun$, thus sampling three adjacent bins in the left panel of Figure~\ref{fig:Percentages1}.
For these horizontally adjacent bins, the percentage of selected SEDs sharply drops from 100\% to 52\% and finally to 0\% with increasing mass. 
The bottom-left panel of Figure~\ref{fig:cc_plots} shows that this trend is due to the models dropping out of the UVX inclusion region (green region with golden border), which selects all sources with $u-g < 0.6$
However, our models with $\mbh= 10^{10}\,\Msun$ have $u-g$ values reaching up to 0.93, and the $10^{10.5}\,\Msun$ models up to 1.39. 
As mentioned in the previous section, the trend with decreasing spin for a given high mass $\geqslant 10^{10}\,\Msun$ (i.e., \emph{vertically} adjacent bins) is similarly driven by the models entering or leaving the UVX inclusion region.

\begin{figure}
\centering
\centerline{\includegraphics[width=0.45\textwidth]{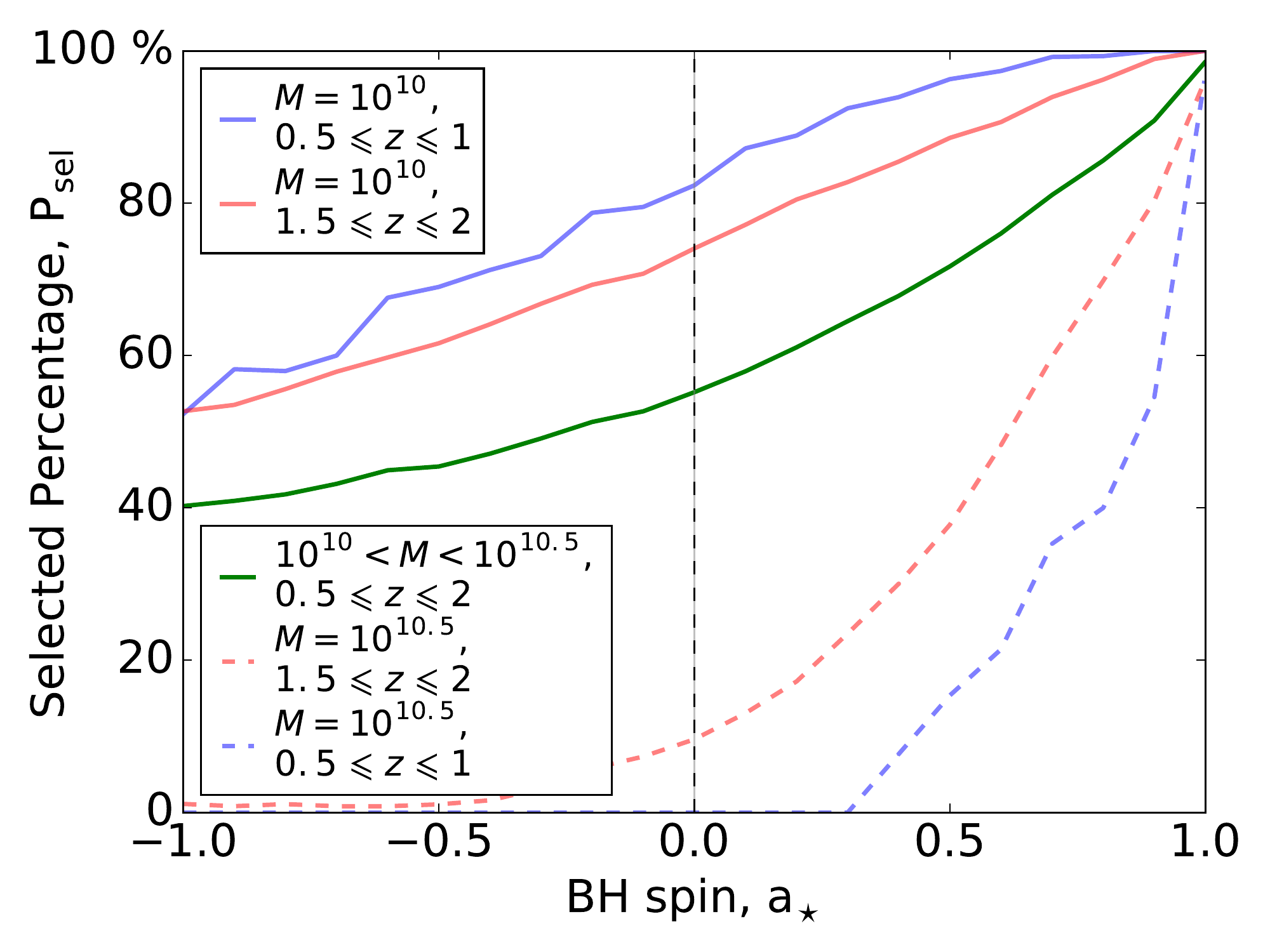}}
\caption{Fraction of selected observable models at a given mass as a function of spin. The blue curves refer to the models at the lower redshift range $0.5 < z < 1$, the red curves refer to those at $1.5 < z < 2$. The solid blue and red curves represent the models with BH mass $\mbh = 10^{10}\, \Msun\ $ and the dotted ones the models with $\mbh=10^{10.5}\, \Msun$. The green curve includes all models with $10^{10} < \mbh <10^{10.5}\, \Msun\ $  for the whole redshift range we simulated, i.e. for $0.5 < z < 2$. 
}
\label{fig:1d_perc}
\end{figure}

In Figure~\ref{fig:1d_perc} we focus on the trends in selection rate with BH spin for high-mass models identified in Figure~\ref{fig:Percentages1}, particularly for models with $\mbh = 10^{10}$ and $10^{10.5}\,\Msun$ in both redshift ranges. 
Fig.~\ref{fig:1d_perc} also displays the fraction of selected models if we include both masses and our \emph{entire} redshift range $0.5 < z < 2$. 
As can be seen, for $\mbh = 10^{10}\, \Msun$, the spin is affecting the higher-redshift models more than the lower-redshift ones, but the opposite is true for $\mbh = 10^{10.5}\, \Msun$. 
In fact, the effect of varying BH spin is most pronounced if the peak of the model SEDs shifts between the spectral regions covered by the $u$ and $g$ bands (which measure the UV excess). 
For the lower mass bin among the two shown in Fig.~\ref{fig:1d_perc}, $\mbh = 10^{10}\, \Msun$, this occurs at higher redshifts, as shown in Fig.~\ref{fig:filter_plot}.
However, for $\mbh = 10^{10.5}\,\Msun$, this occurs at the lower-redshift range since the higher \mbh\ shifts the SED towards longer wavelengths (see Fig.~\ref{fig:DL_reproduced}).

\begin{figure}
\centering
\centerline{\includegraphics[width=0.45\textwidth]{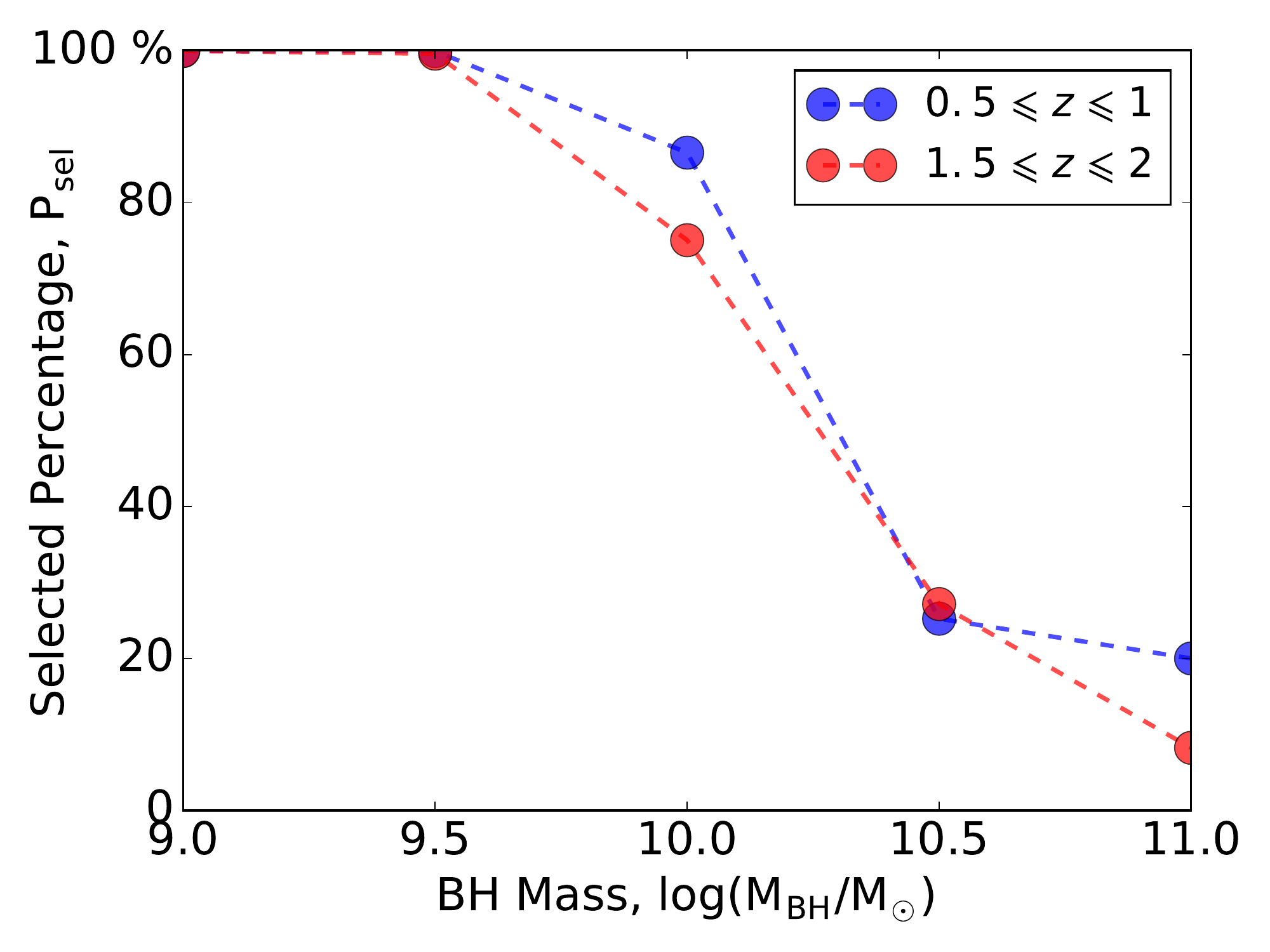}}
\caption{
Fraction of selected observable models as a function of mass. The blue curves refer to the models at the lower redshift range $0.5 < z < 1$, the red curves refer to those at $1.5 < z < 2$.}
\label{fig:1d_perc_fct_of_mass}
\end{figure}

We note that the trend of decreasing selection rate with decreasing BH spin (and/or increasingly retrograde configurations) does not trivially extend to yet higher BH masses.
Returning to Figs.~\ref{fig:Percentages1} and \ref{fig:Percentages2}, we find that 
for models with $\mbh=10^{11}\,\Msun$ at higher redshifts ($1.5 \leqslant z \leqslant 2$), retrograde configurations ($\as<0$) have slightly higher selection rates than most of the prograde ones (see the extreme-right column of bins in the right panels of Figs~ \ref{fig:Percentages1} and \ref{fig:Percentages2}).
This is due to the fact that, unlike the vast majority of the models in our grid, these SEDs are not selected by the UV excess, but instead by SDSS's \textit{griz} criteria, which was designed to select objects at higher redshifts $z \geqslant 2.5$ (see section \ref{sec:SDSSCriteria}). 
Thus, increasing the mass and/or decreasing the spin shifts the AD SED to lower frequencies in a way which is comparable to having a higher redshift. 
However, regardless of spin, the probability of being selected in this high mass regime is still very low, at $<10\%$ (except for the maximum spin value $0.998$, where the rate is $\sim20\%$).
In Figure~\ref{fig:1d_perc_fct_of_mass} we plot the combined probability of selecting a model - that is, summed  over all \as\ values and all redshifts $0.5 < z < 2$, as a function of \mbh.
For $\mbh= 10^{11}\,\Msun$, the selection rate is $\sim$3\%. 
Most of the contribution to this rate comes from models with maximum spin \as $= 0.998$, which have the highest selection rate.
If we exclude the latter, the fraction of selected observable models in this mass regime drops to about 2\%.

Our analysis thus suggests that $z\sim1.5-2$ accreting SMBHs with $\mbh \geqslant 3\times10^{10}\,\Msun$, should they exist, would be virtually impossible to be selected for follow-up SDSS spectroscopy, even if they are accreting at significant fractions of the Eddington limit (i.e, having luminosities that put them above the SDSS flux limit).
Subsequently, such SMBHs would be missing from any demographic analysis that is derived from the SDSS quasar content, which is the largest available (at these redshifts).
As noted above, the possible effects of dust in the nuclei of the host galaxies, and/or of AD outflows, which we did not consider here, would  further challenge the SDSS colour selection, as these would act to decrease the relative contribution of the (rest-frame) UV part of the emergent SEDs \cite[e.g.,][]{Slone2012,Capellupo2015_ADs,Collinson2015_AD}.

Our results add to a number of other biases that work against extremely massive BHs in UV-optical surveys.
It has been recently argued that above a certain threshold BH mass, of roughly $5 \times 10^{10}\,\Msun$ (the exact limit depending on several parameters), it would be impossible to sustain a stable, rotating Keplerian AD outside the ISCO, due to the combined effects of gas and/or radiation pressure \citep{King2016,Inayoshi2016_max_MBH}.
This would imply that accenting SMBHs with such high masses cannot possibly be observed as radiatively-efficient AGN.
Furthermore, it has been suggested that high-mass, low- (or retrograde-) spin BHs
would not be identified by high-ionization emission lines, since the fraction of  ionizing photons of their continuum emission would be too low  \citep{LaorDavis2011_WLQs}.
We also note that \emph{low}-mass nuclear BHs with $\mbh \leqslant 10^{6} \Msun$ may also escape detection, as they may not be able to produce strong broad emission lines \citep[regardless of their intrinsically low luminosities; see][]{Chakravorty2014_lowM_cloudy}.  
Our analysis demonstrates that the difficulty in identifying such objects does not necessarily lie in the existence of some spectral features, but in fact in the practical aspects of simply obtaining a spectrum of an optical source identified in a large, multi-band optical survey. 
From the empirical point of view, recent discoveries of ultraluminous $z>5$ quasars within the SDSS footprint, powered by SMBHs with $\mbh\simeq10^{10}\,\Msun$ \citep{Wang2015_z5_hiM,Wu2015_z6_nature}, have demonstrated the need for non-standard spectroscopic follow-up target selection criteria, which go beyond the standard colour-based procedures.
All this suggests that our current understanding of the high-\mbh, and particularly the low- or retrograde-spin, regime of the SMBH population may be highly incomplete.

\section{Summary and Conclusions}
\label{sec:Summary}

We have tested the completeness of the SDSS quasar colour-based selection algorithm by forward-modelling the spectral energy distributions of accreting, unobscured SMBHs.
This was done by generating a large grid of physically-motivated model SEDs of geometrically-thin, optically-thick accretion discs, varying several key physical parameters. 
In this work we focused on the dominant, accretion disc emission, thus ignoring the more complex emission features seen in real AGN spectra, which are however expected to be of minor importance.
Our models include relativistic corrections and an inclination dependence.
We have obtained the magnitudes of each model SED in the five photometric bands of SDSS, and fed them into the SDSS colour-based quasar selection algorithm. 
Our models overlap with the actual population of colour-selected quasars observed within the SDSS/DR7. 
However, the observed population spans a broader range in colour-colour space than our models.
It is possible that a better match, and therefore a more predictive grid of SEDs, can be obtained by extending the models to include emission lines, observationally-motivated noise, and/or using slim disc models. 
This is however beyond the scope of the present study, 
which focuses on identifying possible biases in the SDSS selection algorithm, particularly with regard to low or retrograde spins among the most massive BHs.

We have found a very high level of completeness across most of the BH mass range, up to $\mbh \simeq 10^{9.5}\,\Msun$: the SDSS colour selection algorithm selects around above 99\% of the model SEDs which satisfy the survey flux limits.
For higher masses, however, we have shown that the completeness drops with increasing mass, and most importantly with decreasing BH spins and for (increasingly) retrograde spin configurations.
Our results imply that if any sources with $\mbh=3\times10^{10.5}$ or $10^{11}\,\Msun$ exist, they would be practically impossible to detect, with an average target identification rate of about 3\%. 
Therefore, large spectroscopic surveys of optically-selected quasars, such as the SDSS, may be missing a significant population of such  objects, even if they do exist as such \cite[see][]{King2016,Inayoshi2016_max_MBH}.
At $\mbh=10^{10}\,\Msun$, we have shown that a maximally retrograde spinning active SMBH would be at most about half as likely to be selected for a spectroscopic follow-up observation as a maximally prograde spinning one. 
This could undermine any attempts to probe SMBH accretion history via spin estimates, which is in principle possible in the extremely high-mass regime. 
Most importantly, our analysis suggests that SMBHs which have grown to extremely high masses through a series of isotropically-oriented accretion episodes, and are thus expected to be ``spun down'', might be missing from large samples of optically-selected quasars.
Thus, some of the recent studies that report high BH spins among the most massive BHs at $z\geqslant1$ \cite[e.g.,][]{Wu2013_eff,Trakhtenbrot2014_hiz_spin,Capellupo2015_ADs,Capellupo2016_XS_pap3} might be affected by selection biases.
We finally note that in this study we only tested the completeness of one key step in the process of quasar detection and identification.
In order to quantify the overall (in-)completeness, one would need to explore possible selection effects in broad and narrow emission line diagnostics. 
Another natural direction for follow-up investigation would be to test to what extent do alternative AGN selection criteria, such as those based on X-ray or radio emission, depend on BH spin.

\section*{Acknowledgements}
We thank the anonymous referee, whose comments helped us to clarify some outstanding issues and improve the presentation of our results.
KS gratefully acknowledges support from Swiss National Science Foundation Grant PP00P2\_138979/1. We thank Andrea Scanzio for providing us with his code recreating SDSS's colour selection algorithm, and for fruitful discussions.
%



\bibliographystyle{mnras}
\bibliography{bibliography2} 





\clearpage

\newpage

\appendix

\section{Lower Eddington ratios}
\label{app:low_lledd}

\begin{figure*}
\centering
\includegraphics[height=0.41\textwidth]{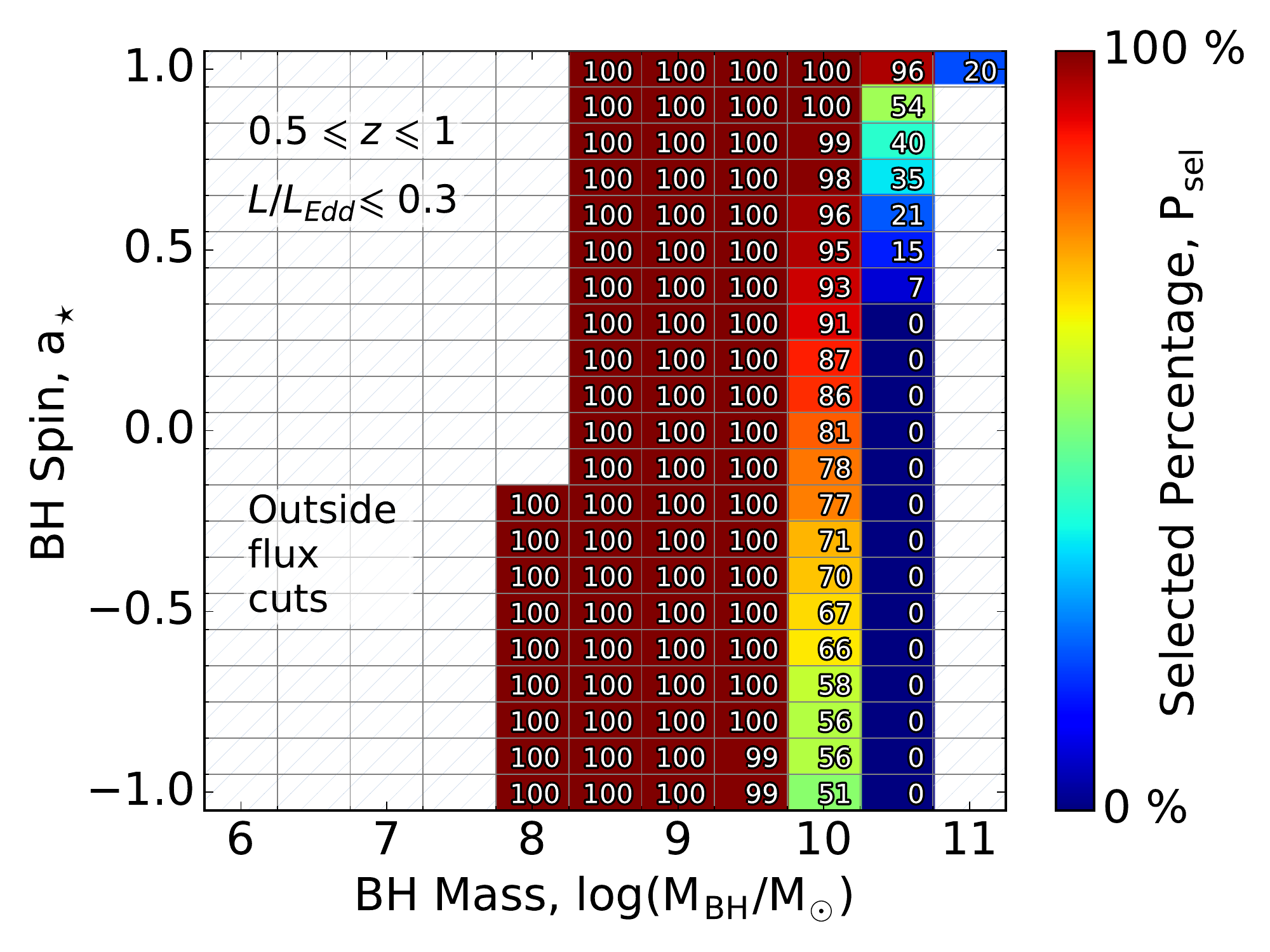}
\includegraphics[trim={0 0 4.1cm 0},clip,height=0.41\textwidth]{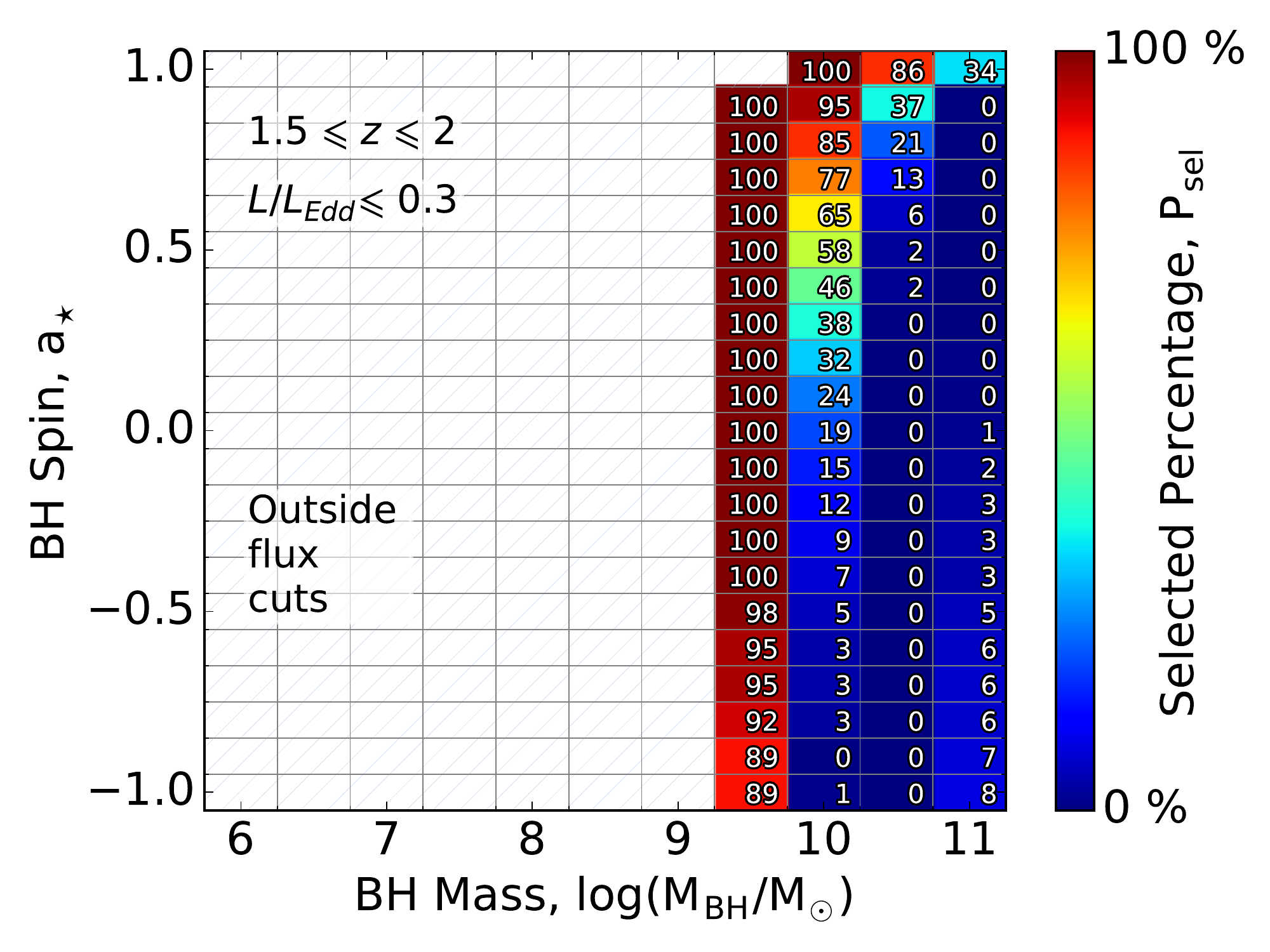} 
\caption{
Same as Figure~\ref{fig:Percentages1}, 
but only for model SEDs with $\lledd \leqslant 0.3$.
} 
\label{fig:Percentages2}
\end{figure*}

\begin{figure*}
\centering
\includegraphics[height=0.41\textwidth]{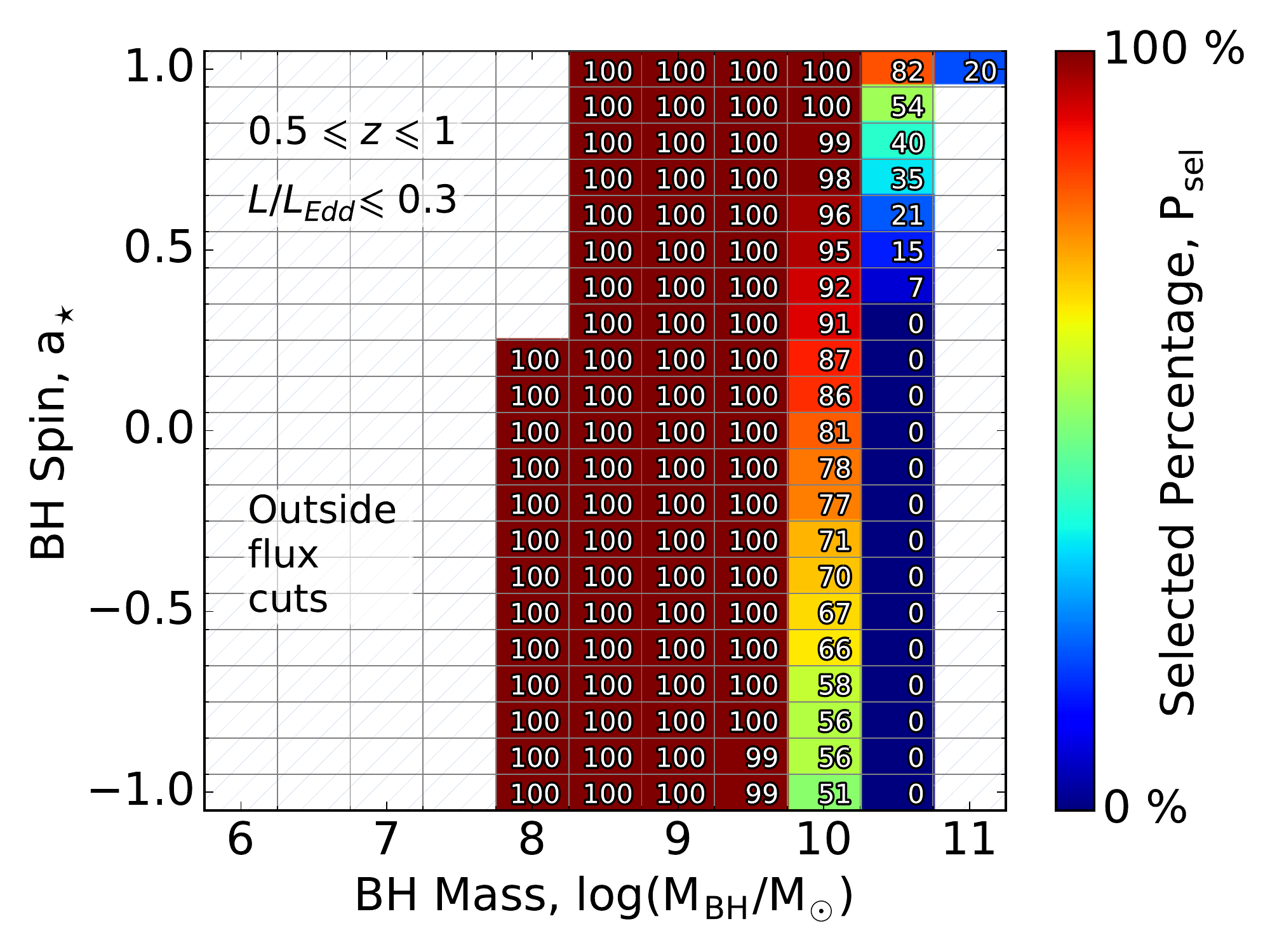}
\includegraphics[trim={0 0 4.1cm 0},clip,height=0.41\textwidth]{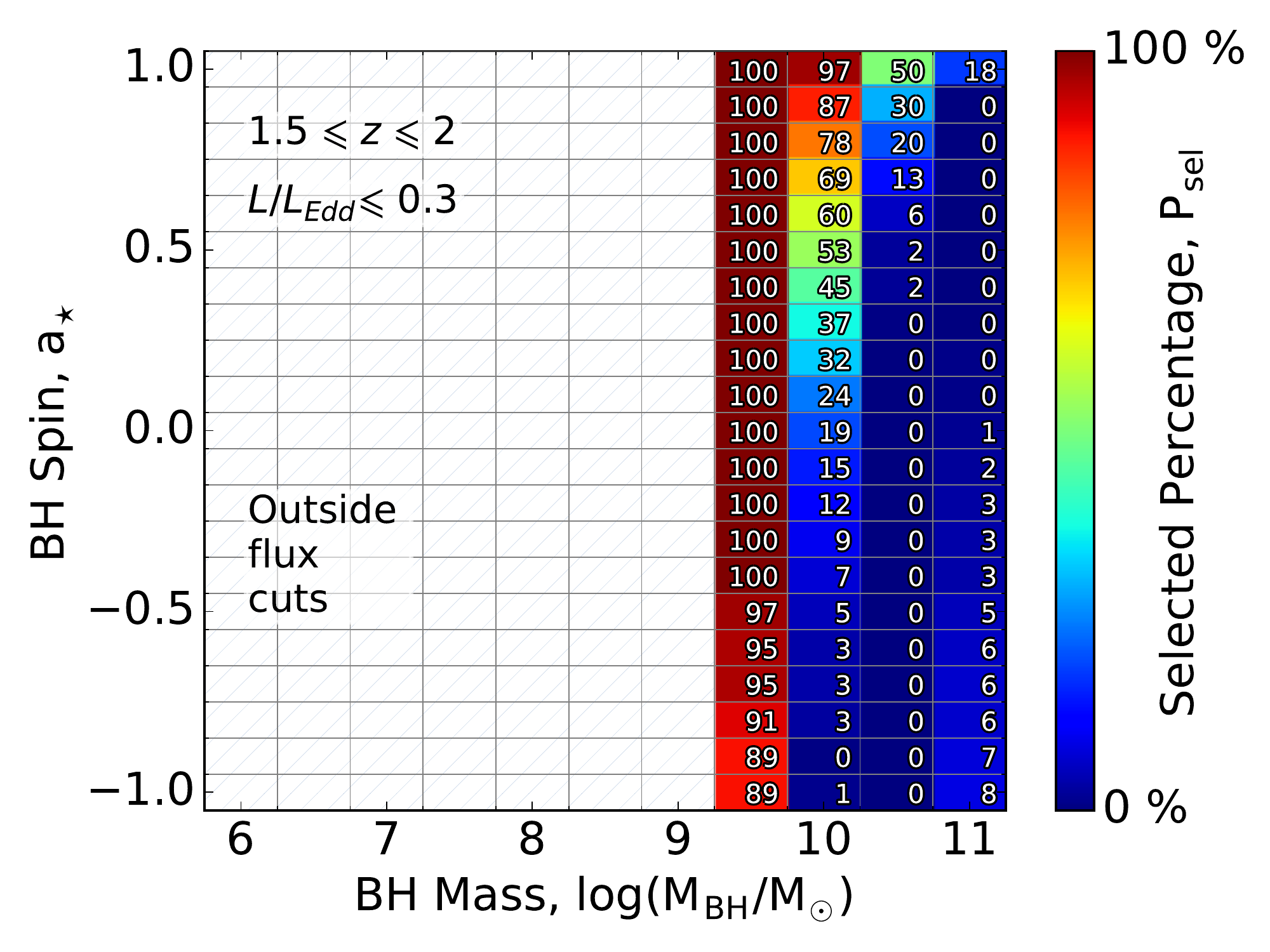} 
\caption{
Same as Figure~\ref{fig:Percentages1_B} (i.e., ignoring temperature colour corrections; $f_{\rm col}=1$), but only for model SEDs with $\lledd \leqslant 0.3$.
} 
\label{fig:Percentages2_B}
\end{figure*}

Here we present alternative versions of Figures~\ref{fig:Percentages1} and \ref{fig:Percentages1_B}, of the percentages of selected model SEDs in the $\mbh-\as$ plane, but accounting only for objects with $\lledd\leqslant0.3$ - the regime where the thin AD model is expected to be valid.
As can be clearly seen in these alternative Figures~\ref{fig:Percentages2} and \ref{fig:Percentages2_B}, the main trends we identified for high-\mbh\ BHs are unchanged. Our main findings regarding a bias of the SDSS colour selection algorithm are therefore not due to our choice of range in \lledd.

\newpage

\section{Testing the Effects of Emission Lines}
\label{app:em_lines_test}

Here we describe the test we performed to verify that our decision to focus on continuum emission, ignoring the emission lines and features, does not significantly affect our findings.

We added a ``typical'' emission line spectrum of unobscured SDSS quasars to all the fiducial calculated AD SEDs (i.e., those \emph{with} temperature colour corrections).
The emission line spectrum was extracted from the composite SDSS quasar spectrum of \cite{VandenBerk2001}.
The composite spectrum was normalized to the best-fitting power-law continuum emission ($f_\nu\propto\nu^{-0.44}$), to produce a \emph{relative} emission line spectrum in the spectral range of $\lambda_{\rm rest}=1190-5100$ \AA\ (the spectral range relevant for our purpose and where the power-law approximation is valid).
Next, each of the fiducial calculated SEDs was multiplied by this relative, composite emission line spectrum.
Similarly to the procedure used for the fiducial SEDs, we then calculated the synthetic magnitudes of these modified SEDs, and followed the SDSS colour-based target selection algorithm.

\begin{figure*}
\centering
\includegraphics[height=0.41\textwidth]{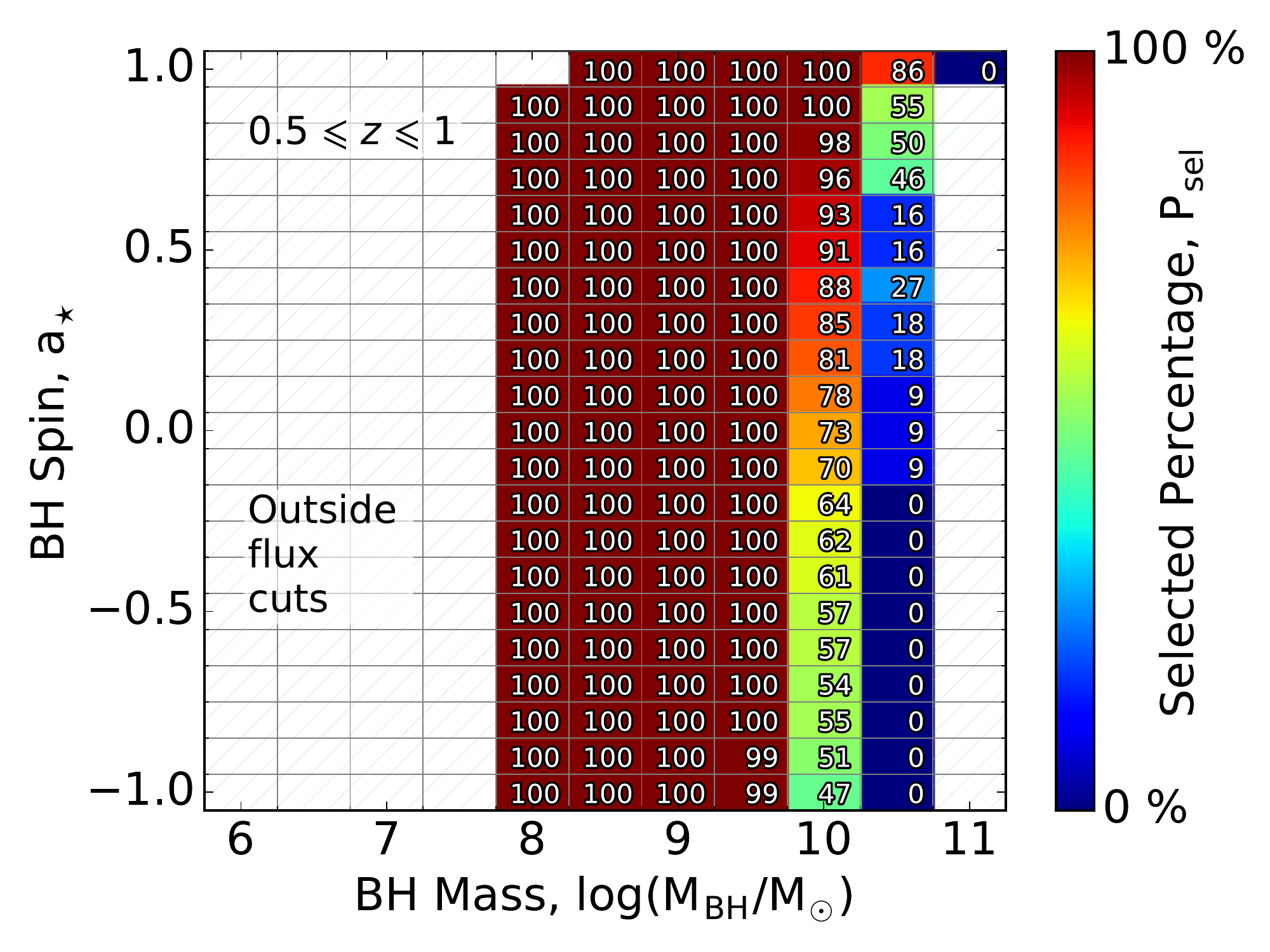}
\includegraphics[trim={0 0 4.1cm 0},clip,height=0.41\textwidth]{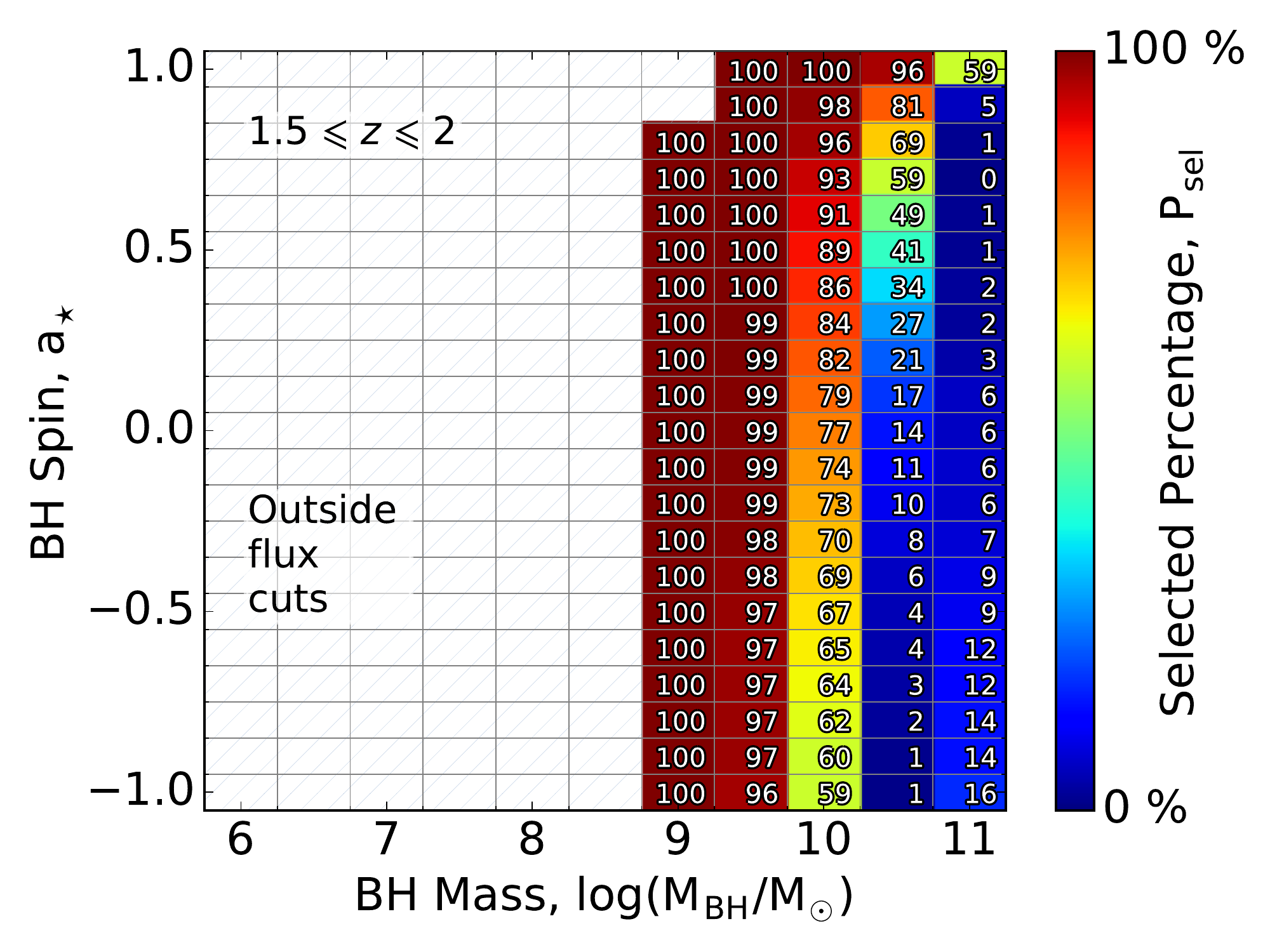} 
\caption{
Same as Figure~\ref{fig:Percentages1}, but for modified model SEDs that include contribution from emission lines and features, typical of SDSS quasars.
} 
\label{fig:Percentages_emlines}
\end{figure*}

Figure~\ref{fig:Percentages_emlines} presents the selection percentages of these modified SEDs in the $\as-\mbh$ plane, similarly to what we show in Fig.~\ref{fig:Percentages1}.
The general trends with BH spin, at the high-\mbh\ regime, remain essentially identical to what we find for the fiducial set of SEDs.
The fraction of SEDs selected by the SDSS algorithm drops with decreasing \as, reaching $\ltsim75\%$ and $\ltsim10\%$ for SEDs with $\as\leq0$, in the mass bins of $\mbh\simeq10^{10}$ and $>10^{10}\,\Msol$, respectively.
A closer inspection of the percentages suggests that these are somewhat higher than those derived for the fiducial model SEDs. 
For example, for maximally retrograde spinning BHs at $z\sim1.8$ with $\mbh\simeq10^{11}\,\Msol$, $\sim15\%$ of the modified SEDs would have been selected by the SDSS algorithm, compared with $\sim7\%$ among the fiducial model.

We stress, however, that the SEDs from such high-\mbh, low-\as\ systems are expected to have extremely weak ionizing radiation fields.
Using SEDs very similar to ours, \cite{LaorDavis2011_WLQs} estimate that for a non-spinning SMBH with $\mbh=10^{10}\,\Msol$ only about $2\times10^{-6}$ of the photons would be ionizing (i.e., beyond the Lyman limit). 
Retrograde spins would result in yet fewer ionizing photons.
Therefore, the strong broad high-ionization lines seen in typical quasars (e.g., \CIV), are expected to be much weaker than what we considered in our test.
This, in turn, would lead to ``redder'' colours, and therefore lower completeness levels in the context of the SDSS colour-based selection procedure.

We conclude that the inclusion of emission lines and features of the strengths observed in typical luminous quasars does not significantly alter our main findings (derived from the fiducial SEDs), and that in reality any such effects are most probably expected to be even more limited.

\bsp	
\label{lastpage}
\end{document}


%% file: mycommands.tex
\newcommand{\todo}{\ifmmode \text{\Huge{\(\bullet\)}} \else {\Huge$\bullet$}\fi}
\newcommand{\tido}{\ifmmode {\bullet} \else $\bullet$\fi}

\newcommand{\E        }[1]{\ifmmode 10^{#1} \else $10^{#1}$\fi}
\newcommand{\tE        }[1]{\ifmmode \times10^{#1} \else $\times10^{#1}$\fi}
\newcommand{\til}{\ifmmode \sim \else $\sim$\fi}
\renewcommand{\~} {\ifmmode \sim \else $\sim$\fi}

\newcommand{\pc}	{\ifmmode {\rm pc} \else pc\fi}
\newcommand{\ld}	{\ifmmode {\rm l.d.} \else l.d.\fi}
\newcommand{\kms}	{\ifmmode {\rm km\,s}^{-1} \else km\,s$^{-1}$\fi}
\newcommand{\cc}	{\ifmmode {\rm cm}^{-3}    \else cm$^{-3}$\fi}
\newcommand{\cmii}	{\ifmmode {\rm cm}^{-2}    \else cm$^{-2}$\fi}
\newcommand{\ergs}	{\ifmmode {\rm erg\,s}^{-1} \else erg s$^{-1}$\fi}
\newcommand{\ergcms}	{\ifmmode {\rm erg\,cm}^{-2}\,{\rm s}^{-1} \else erg\,cm$^{-2}$\,s$^{-1}$\fi}
\newcommand{\ergcmsA}	{\ifmmode {\rm erg\,cm}^{-2}\,{\rm s}^{-1}\,{\rm\AA}^{-1}
\else erg\,cm$^{-2}$\,s$^{-1}$\,\AA$^{-1}$\fi}
\newcommand{  \ergcmsHz  }{\ifmmode{\rm erg\,cm}^{-2}\,{\rm s}^{-1}\,{\rm Hz}^{-1}
                       \else ergs\,cm$^{-2}$\,s$^{-1}$\,Hz$^{-1}$\fi}
\newcommand{\kev}	{\ifmmode {\rm keV} \else keV\fi}

\newcommand{\mic}	{\ifmmode {\rm \mu m} \else $\mu$m\fi}
\newcommand{\vFWHM}	{\ifmmode v_{\mbox{\tiny FWHM}} \else $v_{\mbox{\tiny FWHM}}$\fi}
\newcommand{\vBLR}	{\ifmmode v_{\mbox{\tiny BLR}} \else $v_{\mbox{\tiny BLR}}$\fi}
\newcommand{\sigBLR}	{\ifmmode \sigma_{\mbox{\tiny BLR}} \else $\sigma_{\mbox{\tiny BLR}}$\fi}
\newcommand{\vNLR}	{\ifmmode v_{\mbox{\tiny NLR}} \else $v_{\mbox{\tiny NLR}}$\fi}
\newcommand{\tauBLR}	{\ifmmode \tau_{\mbox{\tiny BLR}} \else $\tau_{\mbox{\tiny BLR}}$\fi}

\newcommand{\Hubble}	{\ifmmode {\rm km\,s}^{-1}\,{\rm Mpc}^{-1} \else km\,s$^{-1}$\,Mpc$^{-1}$\fi}
\newcommand{\NDunit}	{\ifmmode {\rm Mpc}^{-3} \else Mpc$^{-3}$\fi}
\newcommand{\LFunit}	{\ifmmode {\rm Mpc}^{-3}\,{\rm mag}^{-1} \else Mpc$^{-3}$\,mag$^{-1}$\fi}
\newcommand{\MFunit}	{\ifmmode {\rm Mpc}^{-3}\,{\rm dex}^{-1} \else Mpc$^{-3}$\,dex$^{-1}$\fi}

\newcommand{\Msun}{\ifmmode M_{\odot} \else $M_{\odot}$\fi}
\newcommand{\Lsun}{\ifmmode L_{\odot} \else $L_{\odot}$\fi}
\newcommand{\Zsun}{\ifmmode Z_{\odot} \else $Z_{\odot}$\fi}
\newcommand{\mpyr}{\ifmmode \Msun\,{\rm yr}^{-1} \else $\Msun\,{\rm yr}^{-1}$\fi}

\newcommand{\Msol}{\Msun}

\newcommand{\qnote}{\ifmmode q_{0} \else $q_{0}$\fi}
\newcommand{\Hnote}{\ifmmode H_{0} \else $H_{0}$\fi}
\newcommand{\hnote}{\ifmmode h_{0} \else $h_{0}$\fi}
\newcommand{\anote}{\ifmmode a_{0} \else $a_{0}$\fi}
\newcommand{\tnote}{\ifmmode t_{0} \else $t_{0}$\fi}

\newcommand{\ltsim}{\raisebox{-.5ex}{$\;\stackrel{<}{\sim}\;$}}

\def\gsim{\;\rlap{\lower 2.5pt \hbox{$\sim$}}\raise 1.5pt\hbox{$>$}\;}
\def\lsim{\;\rlap{\lower 2.5pt \hbox{$\sim$}}\raise 1.5pt\hbox{$<$}\;}

\newcommand{  \Halpha   }{\ifmmode {\rm H}\alpha \else H$\alpha$\fi}

\newcommand{  \ha       }{\Halpha}
\newcommand{  \Hbeta    }{\ifmmode {\rm H}\beta \else H$\beta$\fi}

\newcommand{  \hb       }{\Hbeta}
\newcommand{  \Hgamma   }{\ifmmode {\rm H}\gamma \else H$\gamma$\fi}
\newcommand{  \Hdelta   }{\ifmmode {\rm H}\delta \else H$\delta$\fi}
\newcommand{  \Lya      }{\ifmmode {\rm Ly}\alpha \else Ly$\alpha$\fi}
\newcommand{  \Lyb      }{\ifmmode {\rm Ly}\beta \else Ly$\beta$\fi}
\newcommand{  \Pa       }{\ifmmode {\rm P}\alpha \else P$\alpha$\fi}
\newcommand{  \Pb       }{\ifmmode {\rm P}\beta \else P$\beta$\fi}
\newcommand{  \Bra      }{\ifmmode {\rm Br}\alpha \else Br$\alpha$\fi}
\newcommand{  \Brg      }{\ifmmode {\rm Br}\gamma \else Br$\gamma$\fi}
\newcommand{  \hii      }{\ifmmode {\rm H}\,\textsc{ii} \else H\,\textsc{ii}\fi}
\newcommand{  \hei      }{\ifmmode {\rm He}\,\textsc{i} \else He\,\textsc{i}\fi}
\newcommand{  \heii     }{\ifmmode {\rm He}\,\textsc{ii} \else He\,\textsc{ii}\fi}
\newcommand{  \HeIIuv   }{\ifmmode {\rm He}\,\textsc{ii}\,\lambda1640 \else He\,\textsc{ii}\,$\lambda1640$\fi}
\newcommand{  \HeIIop   }{\ifmmode {\rm He}\,\textsc{ii}\,\lambda4686 \else He\,\textsc{ii}\,$\lambda4686$\fi}
\newcommand{  \CII	}{\ifmmode \left[{\rm C}\,\textsc{ii}\right]\,\lambda157.74\,\mu{\rm m} \else [C\,{\sc ii}]\ $\lambda157.74\,\mu{\rm m}$\fi}
\newcommand{  \cii	}{\ifmmode \left[{\rm C}\,\textsc{ii}\right] \else [C\,{\sc ii}]\fi}

\newcommand{  \ciii     }{\ifmmode {\rm C}\,\textsc{iii}\right] \else C\,\textsc{iii}]\fi}
\newcommand{  \CIII     }{\ifmmode {\rm C}\,\textsc{iii}\right]\,\lambda1909 \else C\,\textsc{iii}]\,$\lambda1909$\fi}
\newcommand{  \civ      }{\ifmmode {\rm C}\,\textsc{iv}  \else C\,\textsc{iv}\fi}
\newcommand{  \CIV      }{\ifmmode {\rm C}\,\textsc{iv}\,\lambda1549 \else C\,\textsc{iv}\,$\lambda1549$\fi}
\newcommand{  \nii      }{\ifmmode {\rm N}\,\textsc{ii}  \else N\,\textsc{ii}\fi}
\newcommand{  \niii     }{\ifmmode {\rm N}\,\textsc{iii} \else N\,\textsc{iii}\fi}
\newcommand{  \niv      }{\ifmmode {\rm N}\,\textsc{iv}  \else N\,\textsc{iv}\fi}
\newcommand{  \NIVuv    }{\ifmmode {\rm N}\,\textsc{iv}\,\lambda1486 \else N\,\textsc{iv}\,$\lambda1486$\fi}
\newcommand{  \nv       }{\ifmmode {\rm N}\,\textsc{v}   \else N\,\textsc{v}\fi}
\newcommand{\oi}{\ifmmode \left[{\rm O}\,\textsc{i}\right] \else [O\,{\sc i}]\fi}
\newcommand{\OI}{\ifmmode \left[{\rm O}\,\textsc{i}\right]\,\lambda6300 \else [O\,{\sc i}]$\,\lambda6300$\fi}
\newcommand{\oii}{\ifmmode \left[{\rm O}\,\textsc{ii}\right] \else [O\,{\sc ii}]\fi}
\newcommand{\OII}{\ifmmode \left[{\rm O}\,\textsc{ii}\right]\,\lambda3727 \else [O\,{\sc ii}]\,$\lambda3727$\fi}
\newcommand{\oiii}{\ifmmode \left[{\rm O}\,\textsc{iii}\right] \else [O\,{\sc iii}]\fi}
\newcommand{\OIII}{\ifmmode \left[{\rm O}\,\textsc{iii}\right]\,\lambda5007 \else [O\,{\sc iii}]\,$\lambda5007$\fi}
\newcommand{  \OIIIuv   }{\ifmmode {\rm O}\,\textsc{iii}\,\lambda1663 \else O\,\textsc{iii}\,$\lambda1663$\fi}
\newcommand{  \oiv      }{\ifmmode {\rm O}\,\textsc{iv}  \else O\,\textsc{iv}\fi}
\newcommand{  \OIVuv    }{\ifmmode {\rm O}\,\textsc{iv}\,\lambda1402  \else O\,\textsc{iv}\,$\lambda1402$\fi}
\newcommand{  \OIVIR    }{\ifmmode {\rm O}\,\textsc{iv}\,25.9\,\mu {\rm m} \else O\,\textsc{iv}\,$25.9\,\mu$m\fi}
\newcommand{  \ovi      }{\ifmmode {\rm O}\,\textsc{vi}   \else O\,\textsc{vi}\fi}
\newcommand{  \Ovi      }{\ifmmode {\rm O}\,\textsc{vi}\,\lambda1035 \else O\,\textsc{vi}\,$\lambda1035$\fi}
\newcommand{  \nei      }{\ifmmode {\rm Ne}\,\textsc{i}   \else Ne\,\textsc{i}\fi}
\newcommand{  \neii     }{\ifmmode {\rm Ne}\,\textsc{ii}  \else Ne\,\textsc{ii}\fi}
\newcommand{  \NeiiIR   }{\ifmmode {\rm Ne}\,\textsc{ii}\,12.8\,\mu {\rm m} \else Ne\,\textsc{ii}\,$12.8\,\mu$m\fi}
\newcommand{  \neiii    }{\ifmmode {\rm Ne}\,\textsc{iii} \else Ne\,\textsc{iii}\fi}
\newcommand{  \neiv     }{\ifmmode {\rm Ne}\,\textsc{iv}  \else Ne\,\textsc{iv}\fi}
\newcommand{  \nev      }{\ifmmode {\rm Ne}\,\textsc{v}   \else Ne\,\textsc{v}\fi}
\newcommand{  \NevIR    }{\ifmmode {\rm Ne}\,\textsc{v}\,24.3\,\mu {\rm m} \else Ne\,\textsc{v}\,$24.3\,\mu$m\fi}
\newcommand{  \nevi     }{\ifmmode {\rm Ne}\,\textsc{vi}  \else Ne\,\textsc{vi}\fi}
\newcommand{  \mgi      }{\ifmmode {\rm Mg}\,\textsc{i} \else Mg\,\textsc{i}\fi}
\newcommand{  \mgii     }{\ifmmode {\rm Mg}\,\textsc{ii} \else Mg\,\textsc{ii}\fi}
\newcommand{  \MgII     }{\ifmmode {\rm Mg}\,\textsc{ii}\,\lambda2798 \else Mg\,\textsc{ii}\,$\lambda2798$\fi}
\newcommand{  \sii      }{\ifmmode {\rm S}\,\textsc{ii} \else S\,\textsc{ii}\fi}
\newcommand{  \siii     }{\ifmmode {\rm S}\,\textsc{iii} \else S\,\textsc{iii}\fi}
\newcommand{  \siv      }{\ifmmode {\rm S}\,\textsc{iv} \else S\,\textsc{iv}\fi}
\newcommand{  \sili     }{\ifmmode {\rm Si}\,\textsc{i}   \else Si\,\textsc{i}\fi}
\newcommand{  \silii    }{\ifmmode {\rm Si}\,\textsc{ii}  \else Si\,\textsc{ii}\fi}
\newcommand{  \Siliv    }{\ifmmode {\rm Si}\,\textsc{iv}  \else Si\,\textsc{iv}\fi}
\newcommand{  \SilIVuv  }{\ifmmode {\rm Si}\,\textsc{iv}\,\lambda1400  \else Si\,\textsc{iv}\,$\lambda1400$\fi}
\newcommand{  \AlIII   }{\ifmmode {\rm Al}\,\textsc{iii}\,\lambda1857 \else Al\,\textsc{iii}\,$\lambda1857$\fi}
\newcommand{  \Aliii   }{\ifmmode {\rm Al}\,\textsc{iii} \else Al\,\textsc{iii}\fi}
\newcommand{  \caii     }{\ifmmode {\rm Ca}\,\textsc{ii} \else Ca\,\textsc{ii}\fi}
\newcommand{  \feii     }{\ifmmode {\rm Fe}\,\textsc{ii} \else Fe\,\textsc{ii}\fi}
\newcommand{  \feiii    }{\ifmmode {\rm Fe}\,\textsc{iii} \else Fe\,\textsc{iii}\fi}
\newcommand{  \Kalpha   }{\ifmmode {\rm K}\alpha \else K$\alpha$\fi}

\newcommand{ \Lhb   }{\ifmmode L_{\hb} \else $L_{\hb}$\fi}
\newcommand{ \Lha   }{\ifmmode L_{\ha} \else $L_{\ha}$\fi}
\newcommand{ \fwhb  }{\ifmmode {\rm FWHM}\left(\hb\right) \else FWHM(\hb)\fi}
\newcommand{\sighb  }{\ifmmode \sigma\left(\hb\right) \else $\sigma\left(\hb\right)$\fi}
\newcommand{ \ewhb  }{\ifmmode {\rm EW}\left(\hb\right) \else EW(\hb)\fi}
\newcommand{ \fwha  }{\ifmmode {\rm FWHM}\left(\ha\right) \else FWHM(\ha)\fi}
\newcommand{ \ewha  }{\ifmmode {\rm EW}\left(\ha\right) \else EW(\ha)\fi}
\newcommand{ \Lmg   }{\ifmmode L\left(\mgii\right) \else $L\left(\mgii\right)$\fi}
\newcommand{ \fwmg  }{\ifmmode {\rm FWHM}\left(\mgii\right) \else FWHM(\mgii)\fi}
\newcommand{ \Lciv  }{\ifmmode L\left(\civ\right) \else $L\left(\civ\right)$\fi}
\newcommand{ \fwciv }{\ifmmode {\rm FWHM}\left(\civ\right) \else FWHM(\civ)\fi}
\newcommand{ \fwhm  }{\ifmmode {\rm FWHM} \else FWHM\fi} 
\newcommand{ \voff  }{\ifmmode v_{\rm off} \else $v_{\rm off}$\fi} 
\newcommand{ \vmax  }{\ifmmode v_{\rm max} \else $v_{\rm max}$\fi} 

\newcommand{ \mumg  }{\ifmmode \mu\left(\mgii\right) \else $\mu\left(\mgii\right)$\fi}
\newcommand{ \fmg   }{\ifmmode f\left(\mgii\right) \else $f\left(\mgii\right)$\fi}
\newcommand{ \muciv }{\ifmmode \mu\left(\civ\right) \else $\mu\left(\civ\right)$\fi}
\newcommand{ \fciv  }{\ifmmode f\left(\civ\right) \else $f\left(\civ\right)$\fi}

\newcommand{  \auvo     }{\ifmmode \alpha_{\nu,{\rm UVO}} \else $\alpha_{\nu,{\rm UVO}}$\fi}
\newcommand{  \Ledd     }{\ifmmode L_{\rm Edd} \else $L_{\rm Edd}$\fi}
\newcommand{  \lamLlam  }{\ifmmode \lambda L_{\lambda} \else $\lambda L_{\lambda}$\fi}
\newcommand{  \lLl      }{\ifmmode \lambda L_{\lambda} \else $\lambda L_{\lambda}$\fi}
\newcommand{  \nuLnu    }{\ifmmode \nu L_{\nu} \else $\nu L_{\nu}$\fi}
\newcommand{  \nLn      }{\ifmmode \nu L_{\nu} \else $\nu L_{\nu}$\fi}
\newcommand{  \Luv      }{\ifmmode L_{1450} \else $L_{1450}$\fi}
\newcommand{  \Lop      }{\ifmmode L_{5100} \else $L_{5100}$\fi}
\newcommand{  \lLop     }{\ifmmode \log\left(\Lop/\ergs\right) \else $\log\left(\Lop/\ergs\right)$\fi}
\newcommand{  \Lthree   }{\ifmmode L_{3000} \else $L_{3000}$\fi}
\newcommand{  \lLthree  }{\ifmmode \log\left(\Lthree/\ergs\right) \else $\log\left(\Lthree/\ergs\right)$\fi}
\newcommand{  \Lsix      }{\ifmmode L_{6200} \else $L_{6200}$\fi}
\newcommand{  \lLisx     }{\ifmmode \log\left(\Lop/\ergs\right) \else $\log\left(\Lop/\ergs\right)$\fi}
\newcommand{  \Lxray    }{\ifmmode L_{\rm X} \else $L_{\rm X}$\fi}
\newcommand{  \Lhard    }{\ifmmode L_{\rm 2-10} \else $L_{\rm 2-10}$\fi}
\newcommand{  \Lsoft    }{\ifmmode L_{\rm 0.5-2} \else $L_{\rm 0.5-2}$\fi}

\newcommand{\Fthree}{\ifmmode F_{3000} \else $F_{3000}$\fi}
\newcommand{\fuv}{\ifmmode f_{\lambda}\left(1450{\rm \AA}\right) \else $f_{\lambda}\left(1450 {\rm \AA}\right)$\fi}
\newcommand{\fthree}{\ifmmode f_{\lambda}\left(3000{\rm \AA}\right) \else $f_{\lambda}\left(3000{\rm \AA}\right)$\fi}
\newcommand{\fH}{\ifmmode f_{\lambda}\left(1.65\micron\right) \else
$f_{\lambda}\left(1.65\micron\right)$\fi}

\newcommand{\fbol}{\ifmmode f_{\rm bol} \else $f_{\rm bol}$\fi}
\newcommand{\fbolwv}{\ifmmode f_{\rm bol}\left(\lambda\right) \else $f_{\rm bol}\left(\lambda\right)$\fi}
\newcommand{\fbolopt}{\ifmmode f_{\rm bol}\left(5100{\rm \AA}\right) \else $f_{\rm bol}\left(5100{\rm \AA}\right)$\fi}
\newcommand{\fbolthree}{\ifmmode f_{\rm bol}\left(3000{\rm \AA}\right) \else $f_{\rm bol}\left(3000{\rm \AA}\right)$\fi}
\newcommand{\fboluv}{\ifmmode f_{\rm bol}\left(1450{\rm \AA}\right) \else $f_{\rm bol}\left(1450{\rm \AA}\right)$\fi}

\newcommand{\fobs}{\ifmmode f_{\rm obs} \else $f_{\rm obs}$\fi}

\newcommand{  \mbh      }{\ifmmode M_{\rm BH} \else $M_{\rm BH}$\fi}
\newcommand{  \lmbh     }{\ifmmode \log\left(\mbh/\Msun\right) \else $\log\left(\mbh/\Msun\right)$\fi} 
\newcommand{  \lledd    }{\ifmmode L/L_{\rm Edd} \else $L/L_{\rm Edd}$\fi}
\newcommand{  \mmedd    }{\ifmmode \dot{m}/\dot{m}_{\rm \,Edd} \else $\dot{m}/\dot{m}_{\rm \,Edd}$\fi}
\newcommand{  \Lbol     }{\ifmmode L_{\rm bol} \else $L_{\rm bol}$\fi}
\newcommand{  \lbol     }{\ifmmode L_{\rm bol} \else $L_{\rm bol}$\fi}
\newcommand{  \lLbol    }{\ifmmode \log\left(\Lbol/\ergs\right) \else $\log\left(\Lbol/\ergs\right)$\fi} 
\newcommand{  \Lagn     }{\ifmmode L_{\rm AGN} \else $L_{\rm AGN}$\fi}
\newcommand{  \lagn     }{\ifmmode L_{\rm AGN} \else $L_{\rm AGN}$\fi}

\newcommand{  \tgrow     }{\ifmmode t_{\rm growth} \else $t_{\rm growth}$\fi}
\newcommand{  \tUni      }{\ifmmode t_{\rm Universe} \else $t_{\rm Universe}$\fi}

\newcommand{  \Mindot	}{\ifmmode \dot{M}_{\rm infall} \else $\dot{M}_{\rm infall}$\fi}
\newcommand{  \Mbhdot	}{\ifmmode \dot{M}_{\rm BH} \else $\dot{M}_{\rm BH}$\fi}
\newcommand{  \Maddot	}{\ifmmode \dot{M}_{\rm AD} \else $\dot{M}_{\rm AD}$\fi}

\newcommand{  \as	}{\ifmmode a_{\rm *} \else $a_{\rm *}$\fi}
\newcommand{  \avec	}{\ifmmode \vec{a}_{\rm *} \else $\vec{a}_{\rm *}$\fi}
\newcommand{  \re	}{\ifmmode \eta      	 \else $\eta$\fi}
\newcommand{  \RISCO	}{\ifmmode R_{\rm ISCO}  \else $R_{\rm ISCO}$\fi}
\newcommand{  \rg	}{\ifmmode r_{\rm g}  \else $r_{\rm g}$\fi}
\newcommand{  \rS	}{\ifmmode r_{\rm S}  \else $r_{\rm S}$\fi}

\newcommand{  \mseed    }{\ifmmode M_{\rm seed} \else $M_{\rm seed}$\fi}
\newcommand{  \mbul     }{\ifmmode M_{\rm Bulge} \else $M_{\rm Bulge}$\fi} 
\newcommand{  \mstar    }{\ifmmode M_{*} \else $M_{*}$\fi} 
\newcommand{  \mgal     }{\ifmmode M_{*} \else $M_{*}$\fi} 
\newcommand{  \mhost    }{\ifmmode M_{\rm Host} \else $M_{\rm Host}$\fi}
\newcommand{  \mmsmall  }{\ifmmode M_{\rm BH}/M_{*} \else $M_{\rm BH}/M_{*}$\fi}
\newcommand{  \mmlarge  }{\ifmmode M_{*}/M_{\rm BH} \else $M_{*}/M_{\rm BH}$\fi}

\newcommand{  \mmwp     }{\ifmmode \left(M_{*}/M_{\rm BH}\right) \else $\left(M_{*}/M_{\rm BH}\right)$\fi}
\newcommand{  \ml       }{\ifmmode M_{*}/L_{*} \else $M_{*}/L_{*}$\fi}
\newcommand{  \mlwp     }{\ifmmode \left(M_{*}/L\right) \else $\left(M_{*}/L\right)$\fi}
\newcommand{  \mlk      }{\ifmmode \left(M_{*}/L_{K}\right) \else $\left(M_{*}/L_{K}\right)$\fi}
\newcommand{  \sigs     }{\ifmmode \sigma_{*} \else $\sigma_{*}$\fi}
\newcommand{  \Reff     }{\ifmmode R_{\rm e} \else $R_{\rm e}$\fi}
\newcommand{  \nser     }{\ifmmode n_{\rm s} \else $n_{\rm s}$\fi}
\newcommand{  \LFIR     }{\ifmmode L_{\rm FIR} \else $L_{\rm FIR}$\fi}
\newcommand{  \Lfir     }{\ifmmode L_{\rm FIR} \else $L_{\rm FIR}$\fi}

\newcommand{  \mdyn     }{\ifmmode M_{\rm dyn} \else $M_{\rm dyn}$\fi} 
\newcommand{  \mgas     }{\ifmmode M_{\rm gas} \else $M_{\rm gas}$\fi} 
\newcommand{  \mh       }{\ifmmode M_{\rm h} \else $M_{\rm h}$\fi}
\newcommand{  \mhalo    }{\ifmmode M_{\rm halo} \else $M_{\rm halo}$\fi}

\newcommand{  \Lcii     }{\ifmmode L_{\cii} \else $L_{\cii}$\fi}

\newcommand{\bj}{\ifmmode b_{\rm J} \else $b_{\rm J}$\fi}

\newcommand{\iab}{\ifmmode i_{\rm AB} \else $i_{\rm AB}$\fi}

\newcommand{\jab}{\ifmmode J_{\rm AB} \else $J_{\rm AB}$\fi}
\newcommand{\hab}{\ifmmode H_{\rm AB} \else $H_{\rm AB}$\fi}
\newcommand{\kab}{\ifmmode K_{\rm AB} \else $K_{\rm AB}$\fi}

\newcommand{\jveg}{\ifmmode J_{\rm Vega} \else $J_{\rm Vega}$\fi}
\newcommand{\hveg}{\ifmmode H_{\rm Vega} \else $H_{\rm Vega}$\fi}
\newcommand{\kveg}{\ifmmode K_{\rm Vega} \else $K_{\rm Vega}$\fi}

\newcommand{  \Chisq    }{\ifmmode \chi^{2} \else $\chi^{2}$}
\newcommand{  \nelec    }{\ifmmode n_{e} \else $n_{e}$\fi}     
\newcommand{  \nh       }{\ifmmode n_{H} \else $n_{H}$\fi}     
\newcommand{  \Ncol     }{\ifmmode N_{col} \else $N_{col}$\fi} 
\newcommand{  \NH       }{\ifmmode N_{H} \else $N_{H}$\fi}     

\def\deg{\hbox{$^\circ$}}

\def\ion#1#2{#1$\;${\small\rm\@Roman{#2}}\relax}

